\documentclass[aps,physrev,reprint,superscriptaddress,amsmath,amssymb,longbibliography]{revtex4-2}

\usepackage{graphicx}
\usepackage{bm}
\usepackage{microtype}          
\usepackage{xspace}             
\usepackage{booktabs}           
\usepackage{siunitx}            
\usepackage{xcolor}
\usepackage[colorlinks=true,allcolors=blue]{hyperref}
\usepackage{cleveref}           

\graphicspath{{figures/}}

\definecolor{okblue}{HTML}{0072B2}
\definecolor{okorange}{HTML}{E69F00}
\definecolor{okgreen}{HTML}{009E73}
\definecolor{okred}{HTML}{D55E00}

\newcommand{\qsvae}{QSVAE\xspace}
\newcommand{\fsvae}{FSVAE\xspace}
\newcommand{\qlif}{QLIF\xspace}
\newcommand{\maxz}{\ensuremath{\max|z|}}
\newcommand{\auroc}{AUROC\xspace}
\newcommand{\lat}{\ensuremath{\ell}}      

\begin{document}

\title{A quantum-inspired spiking variational autoencoder for robust anomaly detection at the LHC}

\author{Dean Brand}
\email{dean.brand@nithecs.ac.za}
\affiliation{Department of Physics, Stellenbosch University, Stellenbosch, 7600, South Africa}
\affiliation{National Institute for Theoretical and Computational Sciences (NITheCS), Stellenbosch, 7600, South Africa}

\author{Donovan Slabbert}
\affiliation{Department of Physics, Stellenbosch University, Stellenbosch, 7600, South Africa}
\affiliation{National Institute for Theoretical and Computational Sciences (NITheCS), Stellenbosch, 7600, South Africa}

\author{Chanelle Matadah Manfouo}
\affiliation{Department of Physics, Stellenbosch University, Stellenbosch, 7600, South Africa}

\author{Francesco Petruccione}
\affiliation{National Institute for Theoretical and Computational Sciences (NITheCS), Stellenbosch, 7600, South Africa}
\affiliation{School of Data Science and Computational Thinking, Stellenbosch University, Stellenbosch, 7600, South Africa}

\date{\today}

\begin{abstract}
\noindent
Model-agnostic searches for new physics at the Large Hadron Collider utilise unsupervised anomaly detection, where quantum and neuromorphic detectors represent potential approaches. We develop the first fully spiking variational autoencoder, a quantum-inspired spiking variational autoencoder (\qsvae), built on the quantum-inspired leaky integrate-and-fire (\qlif) neuron. In this work we recast the neuron from a measurement-based circuit to a differentiable computational unit. This neuron is benchmarked against other classical, spiking, and quantum models on an LHC di-jet latent space. We show that under noise perturbations the \qlif network retains $0.95$ of its clean accuracy, while accuracy retention for twelve classical models falls to $0.74$--$0.85$. The \qlif anomaly scores respond $3.5$-fold less to the noise perturbation than any of the control model's scores. We compare against a periodic classical control that reproduces the \qlif effect, isolating it to the $\sin^2$ membrane map. This membrane map is a native regulariser that decouples robustness from model capacity supplied by the qubit state space. We demonstrate this through several benchmarks and ablation studies to characterise the performance landscape of the quantum-neuromorphic model space.
\end{abstract}

\maketitle

\section{Introduction}\label{sec:intro}

The search for physics beyond the Standard Model requires methods that do not rely on a predefined signal hypothesis. These model-agnostic approaches do not target a predefined signal but instead train on Standard-Model backgrounds, flagging anything that the detector finds anomalous. This makes it an unsupervised anomaly-detection problem. This problem has resulted in several community benchmarks and a large methodological literature covering reconstruction-based generative models, distance and density estimators, and weakly-supervised classifiers \citep{kasieczka2021lhcolympics,aarrestad2022darkmachines}.

Two computing paradigms have emerged as potential approaches in this setting, each offering possible advantages in efficiency and performance. Quantum machine learning offers kernels based on feature maps which may be difficult to evaluate classically \citep{havlicek2019supervised,schuld2019quantum}, while quantum autoencoders and other quantum anomaly detectors have also been recently proposed for collider data \citep{belis2024quantum,ngairangbam2022quantum,schuhmacher2023unravelling}. Neuromorphic spiking neural networks, on the other hand, offer event-driven and low-power inference on dedicated hardware \citep{davies2018loihi,roy2019towards}, and spiking anomaly detectors have recently been proposed for the latency- and power-constrained environment of the LHC trigger \citep{dillon2026anomaly}. Both are easy to propose but hard to substantiate, because a model can be quantum or biological in inspiration while every operation it performs is actually efficiently classically simulable. Direct comparison against strong classical baselines is therefore important for establishing the relative performance of these approaches on such tasks \citep{han2022adbench}. Recently, other works also report improved adversarial robustness in some spiking neural networks relative to their non-spiking counterparts. Some of these proposed mechanisms include discrete spike encoding, coding choice, membrane leak, and structural parameters such as the firing threshold and time window \citep{sharmin2020inherent,marchisio2020spiking,kim2022rate,elallami2021securing}. Whether a quantum-inspired spiking detector is genuinely competitive, and whether its dynamics add anything beyond that generic spiking robustness, are what we address in this work.

\begin{figure*}[!t]
  \centering
  \includegraphics[width=\linewidth]{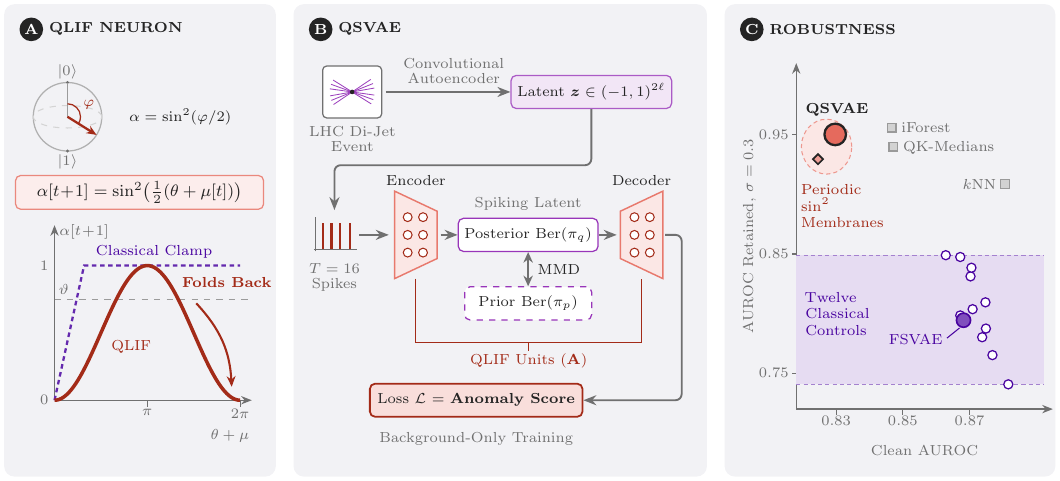}
  \caption{\textbf{Quantum-inspired spiking dynamics decouple noise robustness from model capacity.} \textbf{(A)}~The \qlif neuron carries its membrane potential as a qubit's excited-state population, advanced by one rotation combining the drive $\theta$ with the leak $\mu[t]$, and spikes and resets at the threshold $\vartheta$. The update map is periodic: a strong drive folds the membrane back to 0 where a bounded-classical clamp pins at unity. \textbf{(B)}~The \qsvae places that neuron in every spiking unit of a fully spiking variational autoencoder (\fsvae)~\citep{kamata2022fsvae}, over the LHC di-jet latent space of Belis \textit{et al.}~\citep{belis2024quantum}. A spiking network cannot sample a Gaussian, so the latent code is itself a binary spike train. At each of the $T$ steps, autoregressive spiking networks emit the firing probability of the next latent spike given the spikes so far, and the code is drawn from these Bernoulli outputs. A maximum-mean-discrepancy (MMD) term matches the posterior $\mathrm{Ber}(\pi_q)$ to the prior $\mathrm{Ber}(\pi_p)$, standing in for the intractable Kullback--Leibler divergence. We train on Standard-Model background (or normal) data using a one-class approach, where the per-event loss (reconstruction plus the MMD term) is the anomaly score. \textbf{(C)}~Clean \auroc against the \auroc retained under geometry-respecting noise ($\sigma=0.3$, $\lat=16$). Twelve classical-spiking controls, the \fsvae among them, fall in one band ($0.74$--$0.85$). The shading groups the two families. Note: It is not a confidence region. The two periodic $\sin^{2}$ membranes sit above that band, at the lowest clean accuracy of any spiking model: the \qsvae retains $0.95$, and Periodic (external leak), a control that applies the leak outside the rotation, retains $0.93$. Grey markers are non-spiking references, which are comparably robust, so the separation shown is within the spiking family.}
  \label{fig:feature}
\end{figure*}

We do this with a controlled benchmark built around the \qsvae, the first fully spiking variational autoencoder \citep{kamata2022fsvae} built on the \qlif neuron. We recast that neuron from the Brand--Petruccione \qlif circuit~\citep{brand2024qlif} into a differentiable, gradient-trainable component, and we use it generatively for the first time (\cref{fig:feature}). \qlif is a drop-in spiking nonlinearity with qubit-population membrane dynamics and is classically simulable. Its classical form is a local spiking point-neuron update, making programmable neuromorphic substrates a plausible implementation target \citep{davies2018loihi,roy2019towards}. We make no hardware-realisation claim in this work. We evaluate it together with classical shallow and deep detectors, and with reproductions of the published quantum baselines on the di-jet latent-space data of Belis \textit{et al.}~\citep{belis2024quantum}, which is the benchmark spine that fixes the data, the signals, and the genuinely quantum methods we must implement and match. \Cref{fig:landscape}a summarises the clean-data landscape and \cref{fig:landscape}b previews the central result of this work. Under a fair, geometry-respecting perturbation the \qsvae stays robust despite its low reconstruction capacity, and none of our twelve classical-spiking controls reaches that regime. The controls span the closest alternative explanations: a hard clamp at five firing rates, a smooth-saturating membrane at three settings (default gain, the \qlif map's own slope bound, and matched firing), and the \fsvae with weight decay raised to one hundred times its default. One classical control does reach the regime, the periodic $\sin^2$ membrane with the leak applied outside the rotation, which retains $0.93$ against the \qsvae's $0.95$. The robustness attaches to the $\sin^2$ membrane map, which the qubit state space supplies by its construction. This separation comes with a directly measured $3.5$-fold reduction in how strongly the anomaly scores respond to input perturbation. Throughout this work we report the countervailing evidence: simple statistics win on clean data, isotropic-noise inversion is a noise-model artefact, and a signal-attenuation control shows the robustness to be genuine selectivity.

The rest of this paper is structured as follows, in \Cref{sec:results} we derive and recast the QLIF model as a gradient-trainable neuron and present it in the form of the spiking autoencoder used for anomaly detection. We present its results and benchmarked performance against the other models mentioned above, presenting a case for why the neuron is robust and performs well on this task. In \Cref{sec:discussion} we elaborate on the results presented, highlighting the strengths and limitations of this model as well as the class of task at hand.  In \Cref{sec:methods} we detail the methods used for the benchmarking experiments.


\section{Results}\label{sec:results}

\subsection{The QLIF neuron}\label{sec:model-qlif}

Following Brand and Petruccione~\citep{brand2024qlif}, the \qlif neuron stores its membrane potential as the excited-state population of a single qubit. Writing the qubit state at step $t$ as $\lvert\psi[t]\rangle=\cos(\varphi[t]/2)\lvert0\rangle+\sin(\varphi[t]/2)\lvert1\rangle$, the population
\begin{equation}
  \alpha[t] \;=\; \lvert\langle1\vert\psi[t]\rangle\rvert^2 \;=\; \sin^2\!\big(\varphi[t]/2\big)\in[0,1]
  \label{eq:qlif-mem}
\end{equation}
is the probability of measuring the excited state, which is the quantum analogue of the classical membrane potential. Each step reconstructs the previous step's stored state through the memory rotation $\varphi[t]=2\arcsin\!\sqrt{\alpha[t]}$, where $\varphi[t]\in[0,\pi]$. Two operations then evolve the qubit. Firstly, a weighted input spike rotates it by a trainable angle $\theta$ (integration). Over a delay $\tau$, the excited-state population decays towards zero, similarly as with amplitude damping. This is scaled by $\lambda=e^{-\tau/T_1}$ (where $T_1$ is the relaxation time), and is represented by the decay angle $\gamma[t]=-2\arcsin\!\sqrt{\lambda\,\alpha[t]}$. Dependent on the binary input spike $X[t]\in\{0,1\}$, this create the original update~\citep{brand2024qlif},
\begin{equation}
  \alpha[t+1] \;=\; \sin^2\!\Big(\tfrac{1}{2}(\theta+\varphi[t])\,X[t]\Big)
                 \;+\; \sin^2\!\Big(\tfrac{1}{2}\gamma[t]\,(X[t]-1)\Big),
  \label{eq:qlif-orig}
\end{equation}
where a spike is emitted whenever the population exceeds the firing threshold $\vartheta$, and the qubit reset to $\lvert0\rangle$.

We implement this neuron in a slightly different regime that maintains the core behaviour of the model. Instead of receiving a single binary spike as input, a deep spiking network unit receives a continuous, batch-normalised weighted sum of its pre-synaptic spikes. We take that sum as the rotation drive $\theta$, and apply it at every step with the leak always on, mimicking the behaviour of the classical LIF neuron~\citep{eshraghian_training_2023}. Including the always-on leak in the memory angle through $\mu[t]=2\arcsin\!\sqrt{\lambda\,\alpha[t]}=-\gamma[t]$, integration and leak can be combined into a single rotation,
\begin{equation}
  \alpha[t+1] \;=\; \sin^2\!\Big(\tfrac{1}{2}\big(\theta+\mu[t]\big)\Big),
  \label{eq:qlif}
\end{equation}
which unifies the two branches of \cref{eq:qlif-orig}. With $\theta=0$ it reduces exactly to $\alpha[t+1]=\lambda\,\alpha[t]$ (pure $T_1$ decay), and a non-zero drive rotates the population further. The dynamics depart from the classical LIF neuron in two ways. The membrane is bounded in $[0,1]$, and the non-linearity is the periodic $\sin^2$, so a strong drive rotates the population past threshold and back rather than integrating without bound. As in \cref{eq:qlif-orig}, a spike is emitted and the population reset to zero whenever it crosses the threshold $\vartheta$, before the memory angle of the next step is formed.

The quantum status of this map comes with a caveat. The neuron stores only the excited state population, and the memory rotation re-prepares a pure state from it. The leak reproduces the population dynamics of the amplitude-damping channel within a pure-state ansatz and does not implement the channel itself. True amplitude damping takes a pure state to a mixed one, with coherence $\sqrt{\lambda\alpha(1-\alpha)}$ and $\det\rho=(1-\lambda)\lambda\alpha^{2}>0$. The reconstructed state is pure, with coherence $\sqrt{\lambda\alpha(1-\lambda\alpha)}$, so the scalar memory carries no phase. The two agree exactly on the excited state populations, which is why the zero-drive reduction to $\lambda\,\alpha[t]$ above is exact. Additionally, the reset keeps $\alpha<\vartheta$ when the memory angle is created, which bounds their one-step disagreement in the driven population below $1.1\times10^{-2}$ at these parameters. \Cref{eq:qlif} is therefore the map of a neuron that re-prepares its state from $\alpha$ at every step, as the original circuit does after each measurement, and not the evolution of one qubit persisting across steps. For the same reason, the scalar-$\alpha$ formalism describes independent neurons only. The reduced state of one qubit in an entangled register is generally mixed, and no memory angle can reconstruct it. The reset, by contrast, is exact as a channel statement since taking every state to $\lvert0\rangle$ is amplitude damping at full strength.

In its original form the neuron was a compact circuit, with two rotation gates and no entangling gates, the leakage a rotation reproducing the amplitude-damping population decay, and the spike a measurement, used for forward inference in shallow quantum spiking classifiers. We upgrade it to a trainable building block for deep generative models by reading \cref{eq:qlif} as a differentiable and classically simulable map. We perform backpropagation through the network with a fast-sigmoid surrogate and through the recurrent membrane with the exact leak gradient
\begin{equation}
  \frac{\partial\mu[t]}{\partial\alpha[t]} \;=\; \sqrt{\frac{\lambda}{\alpha[t]\,(1-\lambda\,\alpha[t])}},
  \label{eq:qlif-grad}
\end{equation}
which diverges as $\alpha[t]\to0$. Unrolled across a deep, time-stepped network this destabilises backpropagation-through-time, so we bound the recurrent multiplier in the backward pass. The backward rule also passes a bounded recurrent gradient through the hard reset, whose exact derivative is zero. These backward-pass approximations leave the forward dynamics exactly intact. With this stabilisation the \qlif neuron trains reliably by gradient descent inside an arbitrarily deep spiking network, completing the transition from a single-neuron quantum circuit into a deep-learning component.

It is worth noting that other quantum spiking neurons use different constructions to this approach. Kristensen \textit{et al.} evolve two Hamiltonians and measure locally, so their neuron takes quantum states as both its input and its output \citep{kristensen2021artificial}. Chen \textit{et al.} replace the single-qubit population memory with a multi-qubit stochastic memory and train it by a local learning rule, reporting higher accuracy than the original circuit-based \qlif network on image-classification benchmarks \citep{chen2025stochastic}.

\subsection{The quantum-inspired spiking autoencoder}\label{sec:model-qsvae}

The \qsvae places the \qlif neuron in the fully spiking variational autoencoder of Kamata \textit{et al.}~\citep{kamata2022fsvae}, which is the first variational autoencoder built entirely from spiking layers. We summarise that model here, then describe the substitution process and the adaptation of this model to anomaly detection tasks.

\paragraph{The fully spiking VAE.} In this model, every tensor is a binary spike train over $T$ time steps. A spiking encoder maps the input to spike features in the latent space, and a mirror spiking decoder maps the latent back. The encoder is built from spiking linear layers with threshold-dependent batch normalisation (tdBN)~\citep{zheng2021tdbn} and classical LIF neurons. We use the multilayer-perceptron variant, which consumes a per-event feature vector instead of an image. Sampling a normal variable requires floating-point operations, which a spiking network does not perform, meaning the latent cannot be a continuous Gaussian. Instead is it a Bernoulli spike train produced by an autoregressive spiking model. The posterior $q(z_{1:T}\!\mid x)$ and prior $p(z_{1:T})$ are each autoregressive spiking networks. At each step they emit the probability that the next latent spike will fire, conditioned on the spikes so far, and the latent code is sampled from these Bernoulli outputs. This construction keeps the code binary while making it variational. Based on the standard evidence lower bound of variational autoencoders \citep{kingma2014vae}, the \fsvae replaces the intractable Kullback--Leibler term between spike-train distributions with a maximum-mean-discrepancy (MMD) between samples from the posterior and prior spike trains \citep{kamata2022fsvae},
\begin{equation}
  \mathcal{L} \;=\; \underbrace{\mathbb{E}_{q}\big[-\log p(x\mid z)\big]}_{\text{reconstruction}}
                 \;+\; \underbrace{\mathrm{MMD}\big(q(z),\,p(z)\big)}_{\text{prior matching}}.
  \label{eq:fsvae-loss}
\end{equation}
The reconstruction is read out from the decoder's membrane-potential output accumulated over the $T$ steps, rather than from spikes, making it a real-valued signal comparable to the input.

\paragraph{From FSVAE to QSVAE.} The \qsvae is this network with every LIF neuron replaced by a differentiable \qlif unit. The replacement is exact at the interface because the neuron input and output shapes are identical. Each layer already forms $\theta=\mathrm{tdBN}(\text{weighted input})$, the threshold-dependent batch-normalised pre-synaptic sum, which is the \qlif rotation drive of \cref{eq:qlif}. The neuron does not carry parameters of its own, so the \qsvae has equivalent weight count to the \fsvae, and the latent stays binary, so the model remains fully spiking throughout. Having identically structured networks lets the ablations below neatly attribute effects to the neuron dynamics themselves.

\paragraph{Adaptation to anomaly detection.} We adapt the model from classification to anomaly detection by training it as a one-class model, where it is fitted to Standard-Model background events, learning to reconstruct them and to match the spiking prior. No signal enters the training set. An out-of-distribution event is then flagged by a high per-event loss (\cref{eq:fsvae-loss}, reconstruction plus latent MMD), which is used as the unsupervised anomaly score. To our knowledge this is the first generative, and the first fully spiking variational, use of a quantum-inspired spiking neuron.

\subsection{The clean-data landscape}\label{sec:res-landscape}
We build the benchmark in this work on the di-jet latent space from Belis \textit{et al.}~\citep{belis2024quantum}. A classical convolutional autoencoder compresses each collision event to a per-jet latent vector with dimension $\lat$. A one-class detector is trained on the Standard-Model background (QCD multijet) and scored on a held-out mixture of background and three signals beyond the Standard Model. These signals include: a narrow and a broad Randall--Sundrum graviton decaying to $WW$, and a scalar $A\!\to\!HZ\!\to\!ZZZ$. It should be noted that the detector never sees a signal event during training. In this work we present the per-signal area under the ROC curve (\auroc) together with the inverse background efficiency at fixed signal efficiency, which is the figure of merit used in collider searches.

In this benchmark, fitted detectors genuinely beat model-free statistics, so a comparison among fitted methods is more informative for the model dynamics. At $\lat=16$ the strongest fitted detector reaches an Overall \auroc of $0.891$ (pooled across the three signals; \cref{sec:meth-data}), against $0.791$ for the best model-free statistic. That Overall margin of $0.100$ holds in direction on every signal, but its size varies widely: $0.025$ on the narrow graviton ($0.998$ against $0.973$), $0.108$ on $A\!\to\!HZ$ ($0.987$ against $0.879$), and only $0.022$ on the broad graviton ($0.689$ against $0.667$). The model-free statistics are respectable baselines to compare against while not making the study redundant. This precondition is a property of the regime, and \cref{sec:res-dimension} shows that it weakens as the latent dimension falls.


\begin{figure*}[!htb]
  \centering
  \includegraphics[width=\linewidth]{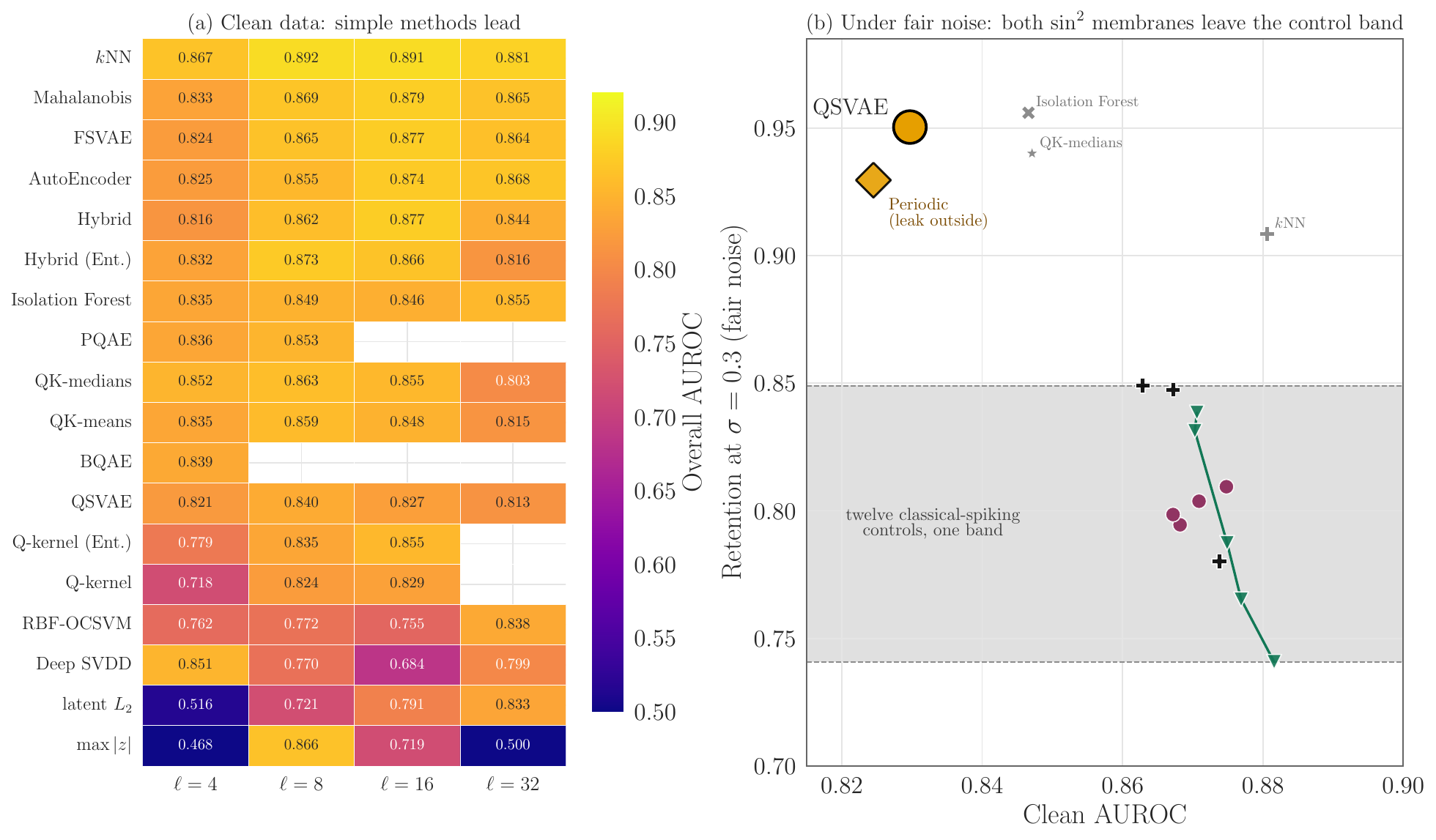}
  \caption{\textbf{Clean-data method comparison and the fair-noise robustness result.} \textbf{(a)}~The clean-data landscape: Overall \auroc for every method (rows, ranked) across the four autoencoder latent dimensionalities $\lat\in\{4,8,16,32\}$ (columns) on the LHC di-jet latent space. Spiking models train on about $13$k background events (a $15$k pool minus a $15\%$ validation holdout), classical methods on about $25$k, the quantum kernel and clustering methods on $600$, and the quantum autoencoders on $400$. Blank cells mark statevector-infeasible configurations: the quantum kernel at $\lat=32$, the persistent quantum autoencoder (PQAE) beyond $\lat=8$, and the bottleneck variant (BQAE) beyond $\lat=4$. The simplest distance method ($k$NN) leads at every dimension and the quantum-inspired \qsvae sits mid-pack: complexity is not rewarded on clean data. \textbf{(b)}~The central result, previewed at $\lat=16$: under a fair, geometry-respecting perturbation of standard deviation $\sigma=0.3$, the twelve classical-spiking control neurons of \cref{sec:res-ablation} all fall in one narrow band of retained \auroc (shaded, $0.74$--$0.85$), while the quantum-inspired \qsvae (large orange circle) and the periodic mechanism control of \cref{sec:res-ablation} (orange diamond) escape it, retaining $0.95$ and $0.93$. Robust reference detectors are shown in grey for context. \Cref{fig:e4-capstone} develops this panel in full.}
  \label{fig:landscape}
\end{figure*}

\begin{table*}[!htb]
  \centering
  \caption{Overall \auroc on the LHC di-jet latent space at $\lat=16$ (representative; per-signal and per-dimension tables and matched-scale rows are in the Supplementary Material, \cref{tab:supp-overall,tab:supp-persignal-l16}). The sixteen rows correspond to fourteen distinct detectors: the quantum kernel and the hybrid autoencoder each appear with and without entanglement. $\sigma_\mathrm{seed}$ is the across-seed standard deviation (five seeds for the spiking models, five for the quantum methods over resampled training pools, three for the quantum autoencoders); the model-free and classical methods are single fits; \cref{tab:supp-search} reports bootstrap $95\%$ confidence intervals over the test set for those with saved per-event scores, in place of a seed spread. $N_\mathrm{train}$ is the approximate training-set size, which differs by family and is disclosed: the model-free statistics involve no fitting, the quantum kernel is limited to ${\sim}600$ events by its $N^2$ Gram matrix, the spiking models train on about $13$k background events, and the classical methods on about $25$k. $1/\varepsilon_b$ is the inverse background efficiency at signal efficiency $\varepsilon_s=0.8$, the search-oriented figure of merit (larger is better).}
  \label{tab:landscape}
  \begin{tabular}{l l c S[table-format=1.3] c S[table-format=1.1]}
    \toprule
    Family & Method & {$N_\mathrm{train}$} & {Overall \auroc $\uparrow$} & {$\sigma_\mathrm{seed}$} &
    {$1/\varepsilon_b@0.8$ $\uparrow$} \\
    \midrule
    Model-free          & \maxz                    & {--}        & 0.719 & -- & 1.9 \\
    Model-free          & L2 energy                & {--}        & 0.791 & -- & 2.2 \\
    \midrule
    Classical (shallow) & $k$NN                    & ${\sim}25$k & \bfseries 0.891 & -- & \bfseries 5.6 \\
    Classical (shallow) & Mahalanobis              & ${\sim}25$k & 0.879 & -- & 4.9 \\
    Classical (shallow) & Isolation Forest         & ${\sim}25$k & 0.846 & -- & 3.2 \\
    Classical (deep)    & AutoEncoder              & ${\sim}25$k & 0.874 & -- & 4.7 \\
    Classical (deep)    & Deep SVDD                & ${\sim}25$k & 0.684 & -- & 1.8 \\
    \midrule
    Spiking             & \fsvae (LIF)             & ${\sim}13$k & 0.877 & 0.005 & 4.4 \\
    Spiking             & \qsvae (\qlif)           & ${\sim}13$k & 0.827 & 0.001 & 2.8 \\
    \midrule
    Quantum             & Quantum kernel (Ent.)    & 600         & 0.855 & 0.003 & 3.7 \\
    Quantum             & Quantum kernel (no ent.) & 600         & 0.829 & 0.004 & 2.8 \\
    Quantum             & QK-medians               & 600         & 0.855 & 0.002 & 3.7 \\
    Quantum             & QK-means                 & 600         & 0.848 & 0.001 & 3.4 \\
    Quantum             & RBF-OCSVM (ref.)         & 600         & 0.755 & 0.034 & 1.8 \\
    Quantum-AE          & Hybrid QAE (non-ent.)    & 400         & 0.877 & 0.008 & 4.4 \\
    Quantum-AE          & Hybrid QAE (Ent.)        & 400         & 0.866 & 0.005 & 4.2 \\
    \bottomrule
  \end{tabular}
\end{table*}

\Cref{fig:landscape}a and \cref{tab:landscape} show sixteen detectors on the clean latent dataset (fourteen in \cref{tab:landscape}; the persistent and bottleneck quantum autoencoders appear only in \cref{fig:landscape}a). The ranking in this is led by the simplest methods: the nearest-neighbour distance ($0.891$) and the Mahalanobis distance to the background ($0.879$) top every group, while the deep one-class baseline (Deep SVDD, $0.684$) sits at the bottom, so complexity is not clearly beneficial in this case. The spiking \fsvae is among the most accurate learned models ($0.877$). It is level with the classical autoencoder and the non-entangled hybrid quantum autoencoder, and is also within $0.015$ of the best distance method. The quantum-inspired \qsvae is in the middle of the pack ($0.827$). The spiking autoencoders are competitive with the genuinely quantum methods. The \fsvae sits above the quantum kernel and the quantum clustering detectors, while the \qsvae sits among them. The cleanest version of this comparison matches the training-set size to the quantum kernel's data budget, which its $N^2$ Gram matrix limits to about $600$ events. At that matched scale the \fsvae ($0.842$) is competitive with the entangled quantum kernel ($0.855$) and even with it on the hardest signal, and both far exceed the RBF one-class support vector machine ($0.755$). No method wins on every signal due to the narrow graviton being easy, with most detectors above $0.99$, while the broad graviton is hard for all (best $0.689$). The pooled Overall score can mask this effect with the \qsvae, for instance, scoring only $0.544$ on the broad graviton (near chance, and below the model-free max $|z|$ statistic at $0.667$), so its Overall figure ($0.827$) rests on the two easier signals. We report per-signal scores throughout (\cref{tab:supp-persignal-l16}), and \cref{sec:res-dimension} returns to the hard signal.

\subsection{Reproducing the quantum baselines}\label{sec:res-quantum}

\begin{figure}[!htb]
  \centering
  \includegraphics[width=0.92\linewidth]{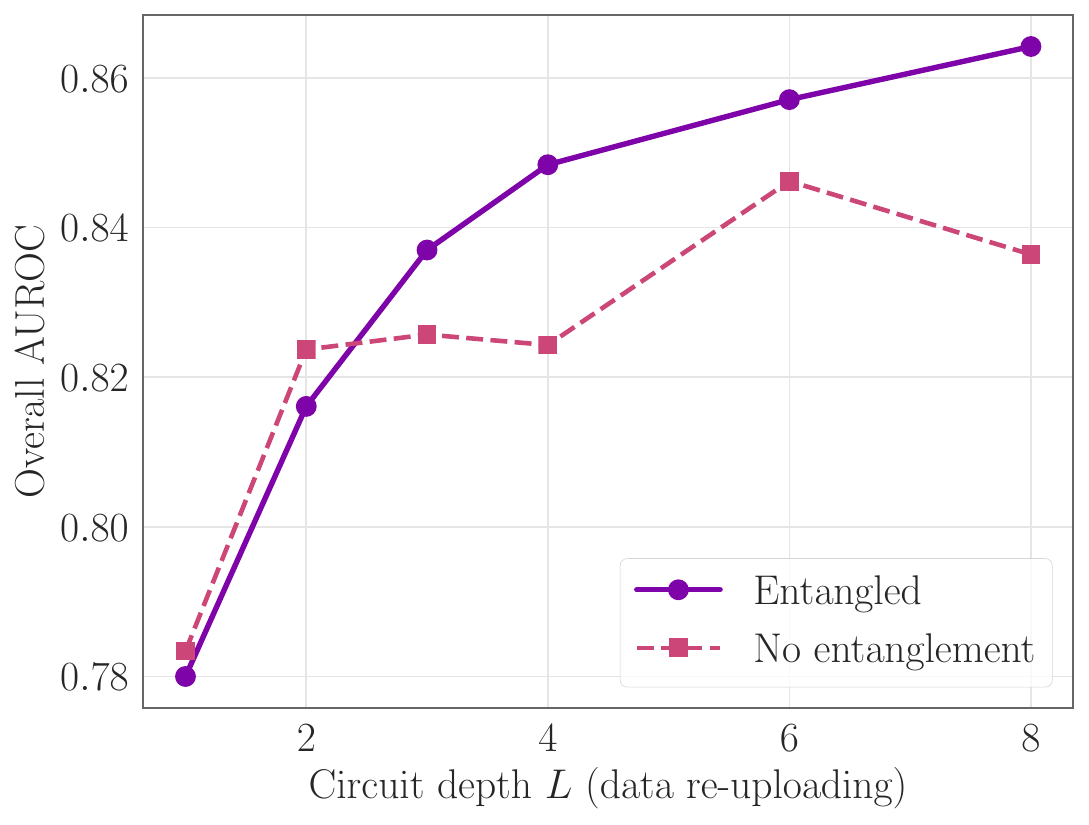}
  \caption{\textbf{Reproducing the quantum-kernel behaviour.} Overall \auroc of the unsupervised quantum kernel machine versus data-reuploading depth $L$ at $\lat=8$, with and without entanglement. Performance and entanglement advantage both grow with circuit depth, reproducing the behaviour reported by Belis \textit{et al.}. Even at maximum depth the quantum kernel approaches but does not reach the best classical method ($k$NN, $0.892$ at $\lat=8$).}
  \label{fig:quantum-reps}
\end{figure}

\begin{figure}[!htb]
  \centering
  \includegraphics[width=0.92\linewidth]{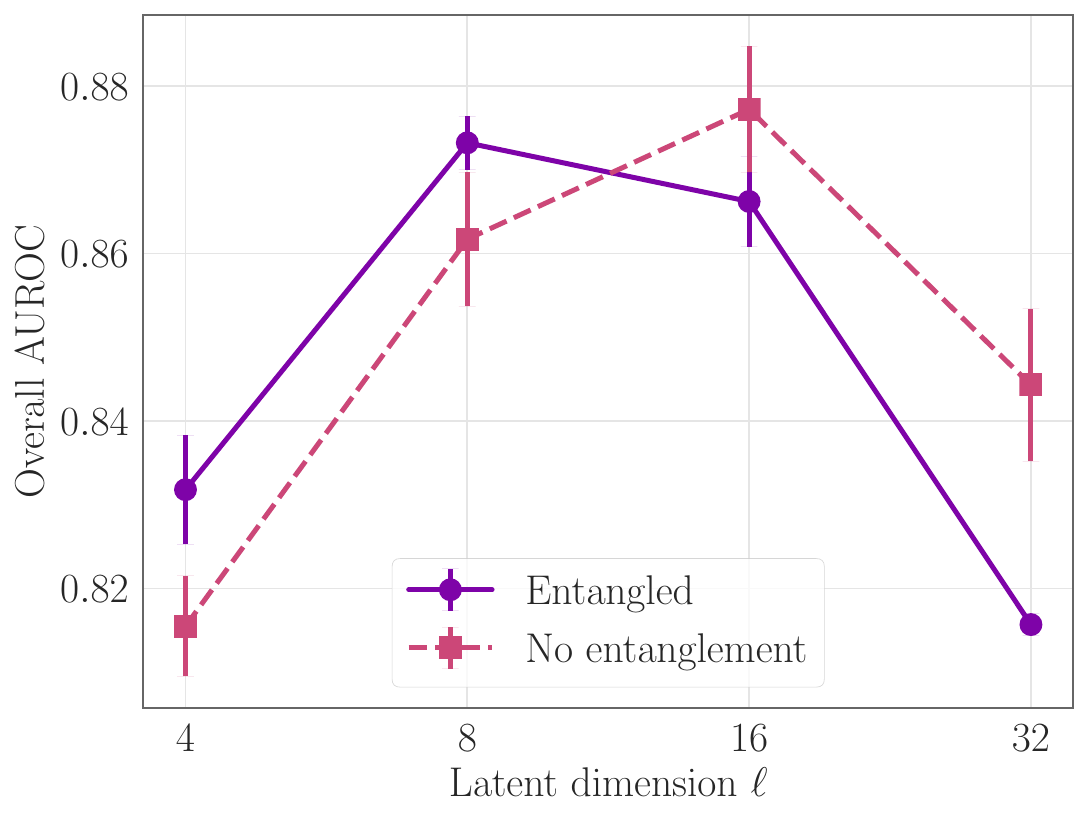}
  \caption{\textbf{The hybrid-QAE entanglement crossover.} Clean Overall \auroc of the hybrid quantum autoencoder (a fixed four-qubit data-reuploading encoder with a classical decoder), with and without entanglement, across latent dimension. The entangled variant peaks at $\lat=8$ and the non-entangled at $\lat=16$. Entanglement helps at low dimension ($\lat\le8$) and hinders at high dimension ($\lat\ge16$), the opposite of the monotone depth advantage of the quantum kernel (\cref{fig:quantum-reps}).}
  \label{fig:hybrid-dim}
\end{figure}

A cross-family comparison for this study is only as fair as its quantum baselines, so we reimplemented the original quantum suite on the same latents and validated its quantum-kernel and RBF-reference subset against the original results. Our reimplementation recovers both of the seed paper's headline findings. The entangled quantum kernel beats its classical RBF reference ($0.855$ against $0.755$ at $\lat=16$). The kernel's performance and its entanglement advantage both grow with data-reuploading depth (\cref{fig:quantum-reps}), which reproduces the reported depth scaling. Against the per-signal \auroc values of Belis \textit{et al.}, our reimplementation of both the quantum kernel and its RBF reference agrees to within $0.065$ across all three signals and both latent dimensions. The largest gap on the near-chance broad graviton where both bracket $0.5$ (\cref{tab:repro-quantum}). Interestingly, in the full comparison, the quantum kernel beats its own classical reference but not the best classical or spiking detectors, and on the hardest signal it is beaten by the $\max|z|$ statistic. We add three quantum autoencoders to the mix as well. The persistent and bottleneck variants \cite{slabbert2026qae} reach Overall \auroc $0.836$ and $0.839$ at $\lat=4$, and the persistent variant reaches $0.853$ at $\lat=8$. They occupy the middle of the clean-data field, while Broad $G$ remains the hardest signal for these models ($0.573$--$0.584$). The single-seed input-noise checks retain $0.812$ (PQAE) and $0.767$ (BQAE) at $\lat=4$, and $0.814$ (PQAE) at $\lat=8$, when purturbed with $\sigma=0.3$. The hybrid variant \cite{manfouo2026hybrid}, which pairs a quantum encoder with a classical decoder, remains feasible across all four widths and leads the quantum-autoencoder family at $\lat=8$ and $\lat=16$ (\cref{fig:hybrid-dim}). It also shows an entanglement crossover where the entangling ring helps at $\lat\le8$ and hinders at $\lat\ge16$, the opposite of the kernel's monotone depth advantage. This is because at high dimension the many re-uploading blocks over-scramble the four-qubit register.

\subsection{Dependence on latent dimension}\label{sec:res-dimension}

\begin{figure}[!htb]
  \centering
  \includegraphics[width=0.92\linewidth]{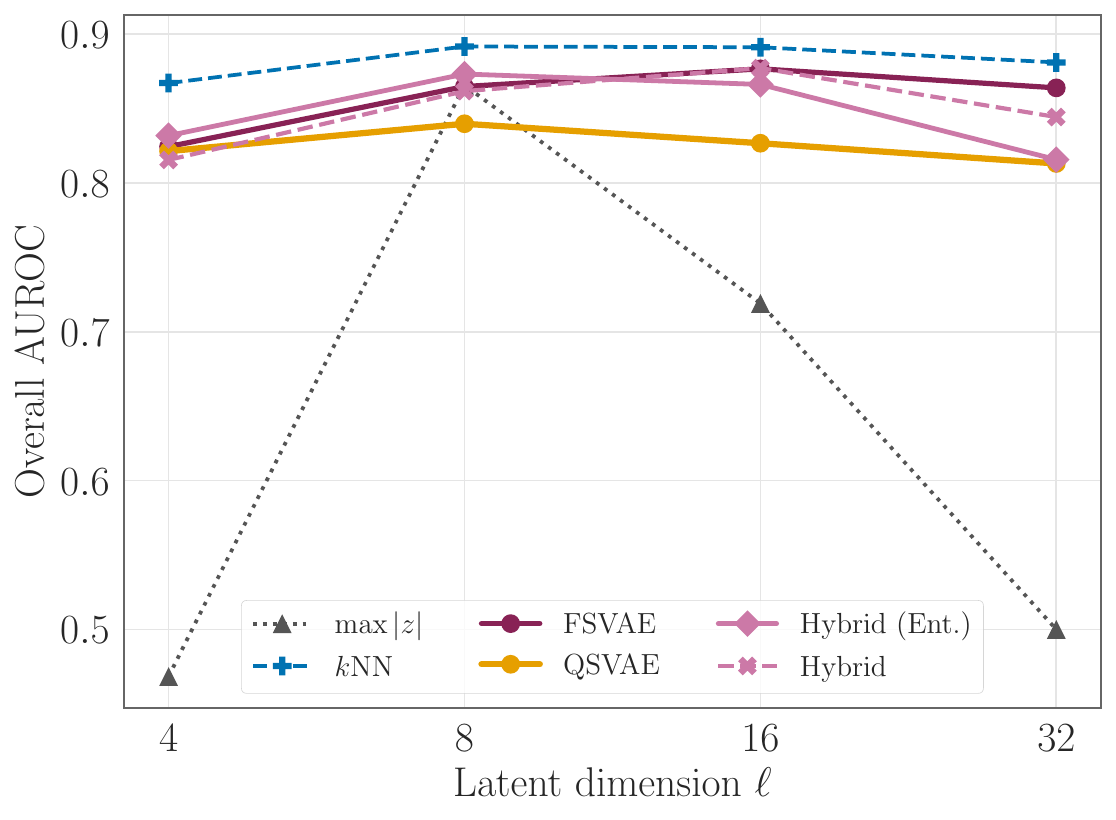}
  \caption{\textbf{Overall \auroc versus latent dimension.} The learned and distance leaders (spiking \fsvae/\qsvae, the best distance method $k$NN, and the hybrid quantum autoencoder, entangled and non-entangled) cluster within a few points of each other at every dimension, while the floor set by the model-free max $|z|$ statistic has a sweet spot at $\lat=8$ and is otherwise weak. The hybrid QAE is competitive with the spiking VAEs and shows its entanglement crossover (\cref{fig:hybrid-dim}).}
  \label{fig:dimension}
\end{figure}

The benchmark's difficulty is not fixed, since the autoencoder latent dimension behaves as a control on it. The model-free statistics expose this effect (\cref{fig:dimension}). The extreme-coordinate statistic max $|z|$ is weak at most dimensions ($0.468$ at $\lat=4$, $0.719$ at $\lat=16$, $0.500$ at $\lat=32$) but has a sharp peak at $\lat=8$ ($0.866$), the dimension at which the quantum baselines are defined and at which quantum hardware would be used. At that peak the model-free statistic ranks fourth of the sixteen detectors, behind only the two distance methods and the hybrid quantum autoencoder. It also ranks ahead of the quantum kernel ($0.835$), both spiking autoencoders, the deep autoencoder, and the isolation forest method. This is because of a geometric difference, in that a tight bottleneck forces an anomalous event to saturate one or more of the $\tanh$-bounded latent coordinates. The extreme-coordinate statistic reads this directly, while a wider bottleneck spreads the signal across coordinates and dilutes it. The learned-versus-simple precondition is therefore a property of the representation rather than the task. The margin between the best distance method and the best model-free statistic shrinks from $0.100$ at $\lat=16$ to $0.026$ at $\lat=8$, which bounds the regime in which any learned detector can justify its complexity.

\subsection{Robustness under a fair perturbation}\label{sec:res-fairnoise}

\begin{figure*}[!htb]
  \centering
  \includegraphics[width=\linewidth]{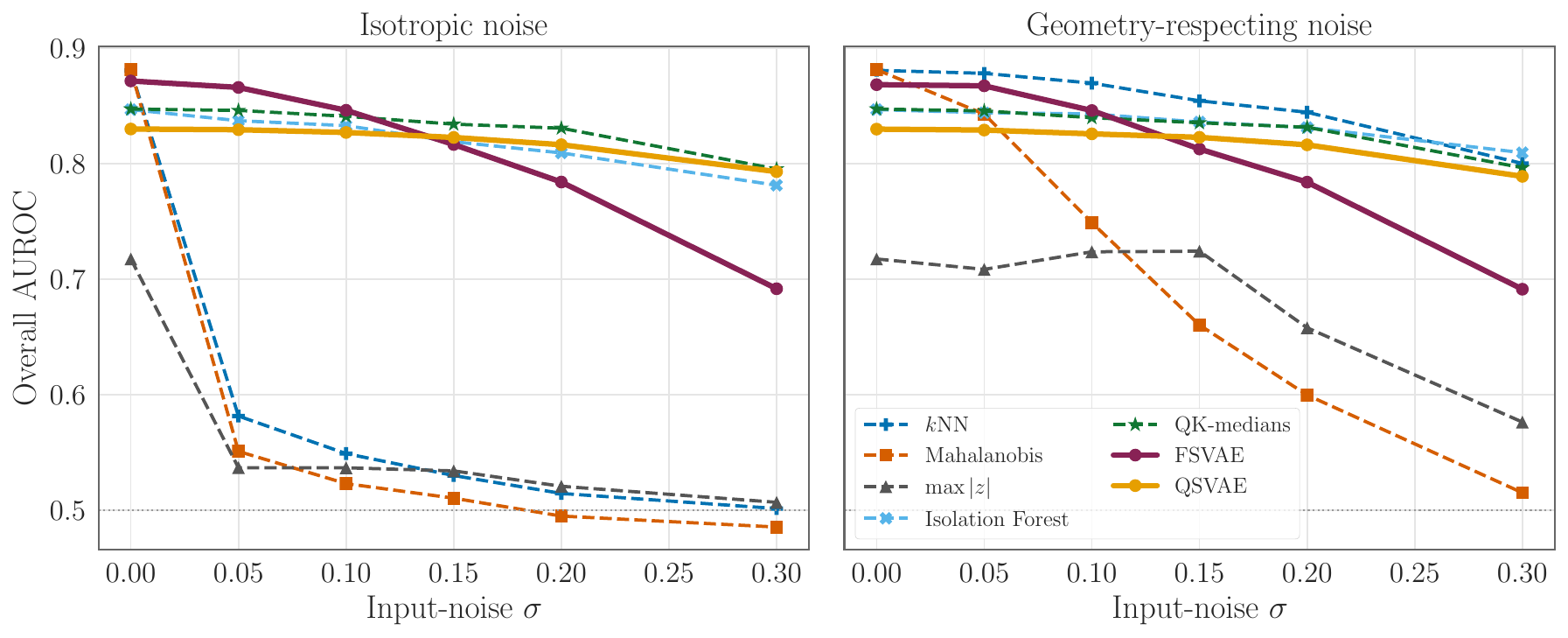}
  \caption{\textbf{Robustness under isotropic and geometry-respecting noise.} Overall \auroc at $\lat=16$ as the test latents are perturbed by additive noise of $\sigma$, every detector fit on clean background and scored on the noised test set. \emph{Left:} isotropic Gaussian noise, which perturbs every latent coordinate equally. The clean-data distance leaders $k$NN and Mahalanobis collapse towards chance. \emph{Right:} geometry-respecting noise, scaled per coordinate by the background's spread and energy-matched so it carries the same total power redistributed to where the data actually varies. The collapse of $k$NN vanishes and it becomes one of the most robust detectors, while Mahalanobis remains brittle. The quantum-inspired \qsvae is among the most robust detectors under both noise models, where the \fsvae is not.}
  \label{fig:e1-fairnoise}
\end{figure*}

A detector trained on simulation data and applied to unseen data must be able to tolerate perturbations (electronic noise, calibration drift, and mismodelling). We fit every method on clean background and then score it on test latents corrupted by additive noise of scale $\sigma$ to test model robustness. We find that the outcome depends sharply on how the noise is shaped. Under isotropic Gaussian noise (\cref{fig:e1-fairnoise}, left), which perturbs every latent coordinate by the same amount, the clean-data leaders collapse. We also see that at $\lat=16$ and $\sigma=0.3$ the nearest-neighbour and Mahalanobis distances fall to $0.501$ and $0.485$, indistinguishable from chance, retaining only $0.57$ and $0.55$ of their clean accuracy. On its own this looks like a clean robustness inversion, with the simplest distance detectors being the most fragile, but this is an artefact of the noise model. Isotropic noise places most of its energy along directions which the background sparsely populates. It disrupts a nearest-neighbour geometry much more than an equal-energy perturbation that is aligned with the data would.

Repeating the test with geometry-respecting noise that is scaled per coordinate by the background's spread and also energy-matched such that it injects the same total power redistributed to where the data actually varies (\cref{sec:meth-noise}), removes the collapse (\cref{fig:e1-fairnoise}, right). At the same perturbation of $\sigma=0.3$ the nearest-neighbour detector now retains $0.91$ of its clean accuracy and is among the most robust methods rather than the least, and the inversion is gone. Mahalanobis is the one exception to this, staying brittle under both models ($0.58$ retained under fair noise) because it scores by the inverse background covariance, amplifying the low-variance directions in which even a spread-matched perturbation is large relative to the background's own scale. The methodological caution we draw from this is that a robustness ranking can be reversed by the choice of perturbation.

Measured against that fair baseline the \qsvae is genuinely robust. It retains $0.95$ of its clean accuracy at $\sigma=0.3$ under geometry-respecting noise ($0.96$ under isotropic noise), making it the most robust of the learned spiking detectors and among the most robust of any kind, on a par with the quantum-clustering methods and just behind the isolation forest ($0.96$ retained). Retention rewards a lower clean baseline, so in absolute terms the \qsvae's robust-regime \auroc ($0.79$) sits within the leading group rather than atop it (isolation forest $0.81$, $k$NN $0.80$, quantum-clustering $0.80$). What matters here is that it far exceeds its architecture-matched spiking twin, the \fsvae ($0.69$, retaining $0.80$), which localises the effect to the neuron itself. The robustness is neither an inversion artefact nor a property the classical-spiking twin shares, which raises the question of what in the \qlif neuron produces it. The next section answers this with a controlled ablation.

\subsection{A firing-matched ablation of the spiking dynamics}\label{sec:res-ablation}

\begin{figure*}[!htb]
  \centering
  \includegraphics[width=\linewidth]{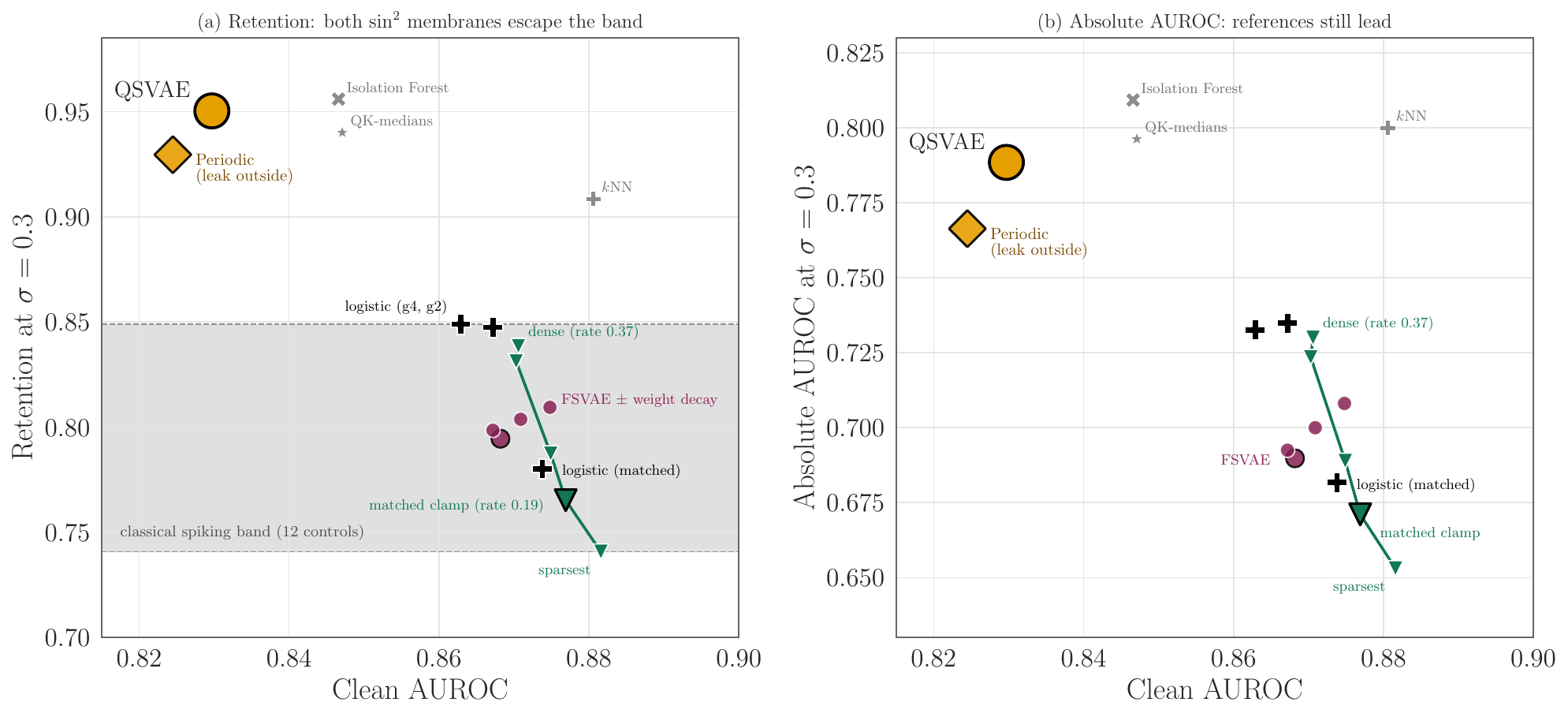}
  \caption{\textbf{Clean accuracy against noise robustness across the spiking family.} Clean \auroc (horizontal) against robustness at a perturbation of $\sigma=0.3$ under geometry-respecting noise, at $\lat=16$. Both panels are windowed to the spiking family of models, with the in-window reference detectors (isolation forest, QK-medians, $k$NN) in grey for context. \textbf{(a)}~Robustness as retention. Twelve classical-spiking controls land in one narrow band (shaded, $0.74$--$0.85$), including the \fsvae at its default and at three higher weight decays (purple circles, default outlined), the clamp swept from sparse to dense firing (green line, firing-matched setting outlined), and the smooth-saturating neuron at its default gain, at the \qlif slope bound, and firing-matched (black crosses). Two variants break out of the band, at the lowest clean \auroc of any spiking model: the quantum-inspired \qsvae (orange circle), which retains $0.95$, and the Periodic (external leak) control (orange diamond), the $\sin^2$ membrane with the leak moved outside the rotation term, which retains $0.93$. \textbf{(b)}~The same variants in absolute \auroc under noise: the two periodic membranes lead the spiking family, while the strong references lie at or slightly above it (the Mahalanobis and max $|z|$ references fall below the plotted window), so the distinctive grouping is relative to the spiking family specifically, rather than to all detectors.}
  \label{fig:e4-capstone}
\end{figure*}

\begin{figure*}[!htb]
  \centering
  \includegraphics[width=\linewidth]{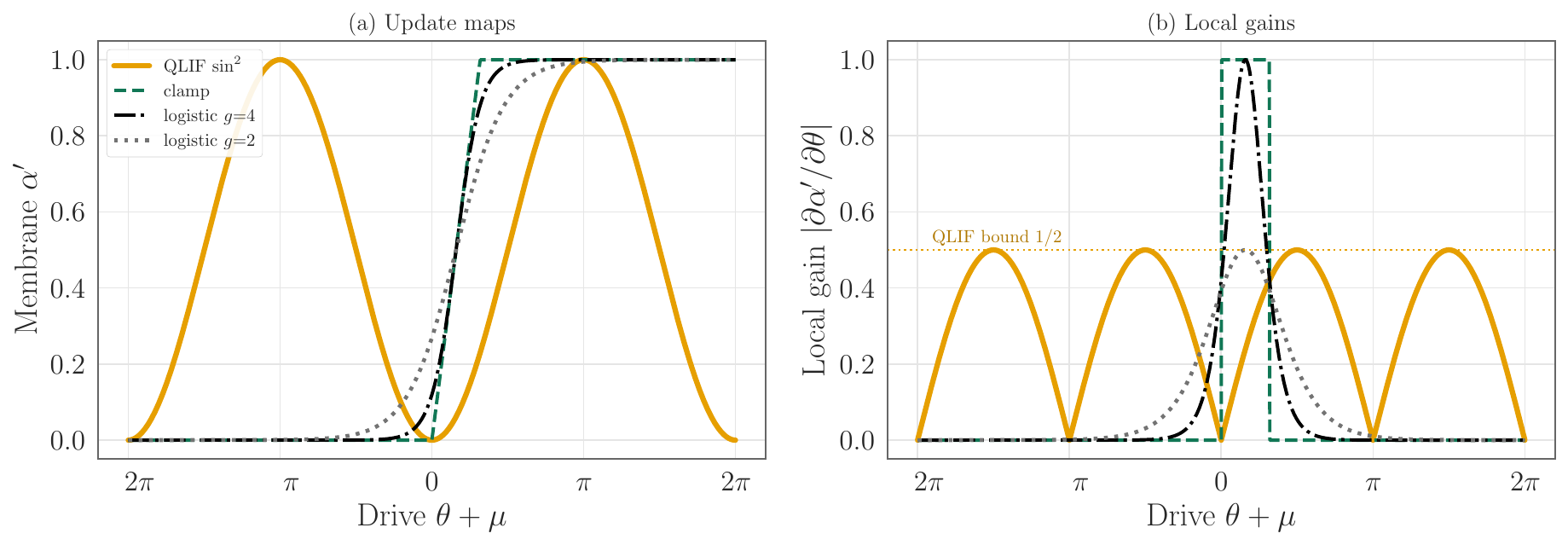}
  \caption{\textbf{The \qlif update map against classical control variants.} \textbf{(a)}~The membrane update $\alpha'$ as a function of the input drive, for the \qlif $\sin^2$ map of \cref{eq:qlif} and three bounded-classical membranes used as ablation controls as follows: the hard clamp, and the smooth-saturating (logistic) neuron at gains $4$ and $2$. All four are bounded to the unit interval and three of them saturate smoothly, but only the $\sin^2$ map is periodic with a strong drive folding the membrane back instead of pinning it. \textbf{(b)}~Corresponding local gains. The \qlif gain never exceeds one half (dotted line), the clamp has unit gain throughout its interior, and the smooth-saturating neuron reaches $g/4$ at its centre, so the $g=2$ setting matches the \qlif bound. These maps organise the ablation: boundedness, smoothness, slope, and firing are each reproduced by a control, and periodicity is what remains unique to the \qlif neuron.}
  \label{fig:membranes}
\end{figure*}

The \qsvae differs from the \fsvae only in its neuron, so the robustness of \cref{sec:res-fairnoise} is due to a property of the \qlif neuron. Its bounded $[0,1]$ membrane, its sparse firing, and its $\sin^2$ rotation (\cref{eq:qlif}) are three candidate properties, which we separate with a family of classical-spiking controls. Each of these controls is trained identically to the \qsvae except for the neuron (\cref{fig:membranes} shows the competing update maps). The Clamped neuron has the \qlif neuron's bounded $[0,1]$ membrane, threshold, and leak, but replaces the $\sin^2$ rotation with linear leaky integration and a hard clamp so it is bounded exactly as the \qlif neuron is, but fully classical. Because a sparser neuron is mechanically more robust, we sweep the clamp across five firing rates (from about $0.2$ to $0.37$ at the settings probed). One of them (Clamped (matched)) is tuned close to the \qsvae's rate ($0.19$ against $0.23$). Smooth saturation is the next candidate, since a membrane whose sensitivity vanishes towards its bounds is the property one might credit for the $\sin^2$ map. To evaluate this we add a smooth-saturating neuron whose bounded membrane follows a logistic curve rather than the periodic $\sin^2$,. It is trained at three settings including its default gain, the gain at which its maximum slope matches the \qlif map's global bound of one half, and firing-matched to the \qsvae. To ask whether ordinary regularisation would reach the same regime, we train the \fsvae at three weight-decay strengths up to $100$ times the default. Between them these twelve controls vary boundedness, firing density, saturation smoothness and slope, and regularisation strength independently of the $\sin^2$ dynamics. We find that they do not vary heterogeneity, as every variant, including the \qlif, uses one fixed leak ($T_1=60$ steps) for every neuron. The diversity in per-neuron time constants known to improve the robustness of spiking networks \citep{pereznieves2021neural} is available to none of them, including the \qsvae.

The controls overturn the simple ``bounded membrane'' explanation of the accuracy. Boundedness is free, as the Clamped neuron matches the \fsvae's clean \auroc ($0.876$ against $0.877$ at $\lat=16$, five-seed), and the smooth-saturating neuron likewise ($0.876$). At matched firing the bounded-classical neuron is the most accurate of every variant ($0.883$), all comfortably above the \qsvae ($0.827$). The \qsvae's clean-accuracy cost is therefore specific to the $\sin^2$ dynamics, and attaches to neither the bounded membrane nor the firing rate.

For the robustness analysis, the controls isolate the effect (\cref{fig:e4-capstone} and \cref{tab:supp-controls}). Every classical-spiking variant we trained falls in a single narrow retention band, between $0.74$ and $0.85$. In the band there are twelve variants in total, including the \fsvae, the clamp at all five firing rates, the smooth-saturating neuron at all three settings, and the \fsvae at all three regularisation strengths. The \qsvae breaks out of the band, retaining $0.95$. Three comparisons narrow down what carries the difference.

Firing density does not explain the above. The clamp traces a monotone frontier as its rate varies, sparse firing is more accurate and less robust (down to $0.74$ retained), and dense firing is more robust (up to $0.84$). No firing rate brings the clamp near $0.95$, and the firing-matched clamp, if anything more sparse than the \qsvae, retains only $0.76$. Smooth saturation does no better. The smooth-saturating neuron has exactly the vanishing-derivative-at-saturation property one might credit for the robustness, yet it lands on the clamp ($0.85$). Matching its maximum slope to the \qlif bound of one half leaves it there ($0.85$), and firing-matching it moves it down the classical frontier ($0.78$). This is the same direction the matched clamp moved, so it does not suffice at any slope or firing rate that we tested. Ordinary regularisation does not reproduce it either, as scaling the weight decay by up to $100$ times leaves the \fsvae at $0.80$. Across firing density, saturation shape, and regularisation strength, none of the twelve classical-spiking controls reaches the \qsvae's corner, so the robustness is attributed to the $\sin^2$ dynamics.

A further control tests if the leak must be inside the memory angle. We train a periodic membrane, Periodic (external leak), that keeps the $\sin^2$ rotation but applies the leak as an external multiplicative damping factor instead of folding it into the memory angle. It breaks out of the band with the \qsvae, retaining $0.93$ at $\lat=16$ and $0.91$ at $\lat=8$, and it pays the same clean-accuracy cost ($0.822$ against the \qsvae's $0.827$). The $\sin^2$ membrane map therefore carries the effect, and embedding the leak inside the rotation adds only a small further margin ($0.95$ against $0.93$). Both escaping variants share the full map, its periodic fold-back together with the pure-state memory reconstruction that generates it, so this control does not separate those components from each other. The controls also distinguish the effect from the robustness that has been attributed to spiking networks in general \citep{sharmin2020inherent,elallami2021securing}. Every control shares the \qsvae's discrete spike encoding and leak, and the firing sweep spans its sparsity, but the gap remains.

We further examine the robustness advantage of the \qsvae through three additional checks. The separation is not an artefact of pooling the three signals, as we find the \qsvae retains $0.97$, $0.97$, and $0.90$ on the narrow graviton, the broad graviton, and $A\!\to\!HZ$ respectively, above the control band on each. On the broad graviton, however, that retained value sits on a near-chance clean \auroc and so has little significance. Retention also rewards a low clean baseline, so among detectors of all families the isolation forest, $k$NN, and the quantum-clustering methods are comparably robust (\cref{sec:res-fairnoise}). It holds in absolute robust \auroc as well, with $0.79$ for the \qsvae against $0.73$ to $0.65$ for the spiking controls. The core ordering also holds at $\lat=8$, where the \qsvae retains $0.91$ against $0.81$ to $0.85$ for the default clamp, the firing-matched clamp, the default smooth-saturating neuron, and the \fsvae. The full firing and regularisation sweeps were run only at $\lat=16$.

The training stability points in the same direction, albeit more weakly. At $\lat=16$ the \qsvae's across-seed standard deviation ($0.0012$) is roughly four times smaller than the \fsvae's ($0.0048$) and the dense Clamped neuron's ($0.0051$). Firing-matching the clamp recovers part of it ($0.0028$) but leaves the \qsvae more than twice as stable. This advantage is specific to $\lat=16$, and at $\lat=8$ no spiking variant is reliably steadier, so it cannot be claimed as a general property.

To investigate the source of this robustness, we measure the models' sensitivity to input perturbations directly. For each trained model we perturb the test inputs by a small geometry-respecting step and record how much the anomaly score moves, relative to the score's own spread (\cref{sec:meth-noise}). The \qsvae's scores are $3.5$-fold less sensitive to this perturbation than the least sensitive of the twelve classical-spiking controls', shown by a normalised sensitivity of $0.17$ for the \qsvae against $0.60$ to $0.70$ across the full control family. This separation holds at every probe scale we tested ($\sigma\in\{0.05,0.1,0.2\}$). A normalisation-free readout confirms it is the score ordering, the quantity \auroc depends on, that is preserved. Under the same perturbation the \qsvae's score ranking barely moves while every probed control's scrambles ($\tau=0.40$ to $0.51$; $17$ to $20\%$). The same tests localise the effect to the dynamics, where the firing-matched clamp, which is the sparsest of the probed variants, is if anything the most sensitive ($0.70\pm0.02$. This happens at the top of the $0.61$--$0.70$ control range). The smooth-saturating neuron is as sensitive as the hard clamp, so neither firing density nor smooth saturation accounts for it. The bounded $\sin^2$ membrane also settles at a reconstruction floor about four times the classical neuron's (\cref{fig:reconfloor}) without depressing accuracy. This is consistent with a dynamics-specific regularisation of the code. Whether that compression is itself the route to robustness we leave open.

\subsection{A signal-attenuation control}\label{sec:res-sensitivity}

\begin{figure}[!htb]
  \centering
  \includegraphics[width=\linewidth]{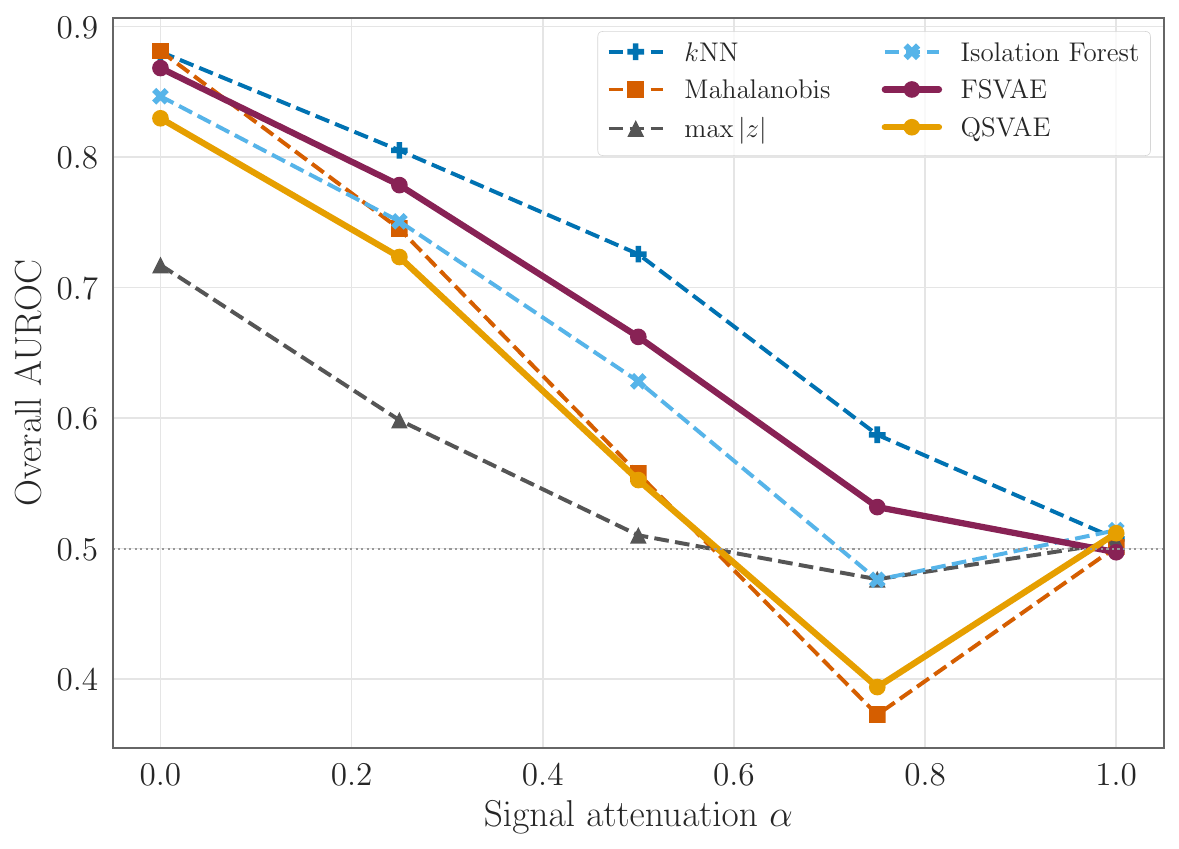}
  \caption{\textbf{AUROC under signal attenuation at $\lat=16$.} Overall \auroc at $\lat=16$ as the test signal is attenuated towards the background, from the true signal ($\alpha=0$) to a ``signal'' statistically identical to background ($\alpha=1$). Every detector, including the \qsvae, loses its separation as the discriminating structure is removed, converging to chance at $\alpha=1$. The intermediate dip below $0.5$, shared with the Mahalanobis and isolation-forest scores, shows partly-morphed signal events sitting deeper in the background than typical background signals. A detector insensitive to its input would instead stay flat.}
  \label{fig:e2-sensitivity}
\end{figure}

A detector whose scores barely move under input noise (\cref{sec:res-fairnoise}) is tempting to read as deflationary, as perhaps the \qsvae is insensitive because it discards information, assigning near-constant scores that happen to resist perturbation. If that were the case, its robustness would be a degeneracy. We find evidence against this explanation through a signal-attenuation control that runs the opposite test to the noise sweep, where the noise sweep corrupts the input, this control removes the signal. We interpolate each test-signal event towards the background, scoring at attenuations $\alpha$ from the true signal ($\alpha=0$) to a signal that is statistically identical to background ($\alpha=1$) (\cref{sec:meth-noise}). A detector that genuinely reads the discriminating structure must lose its separation as $\alpha\to1$, as a detector that ignores its input would not.

The \qsvae behaves as a signal-tracking detector is intended to, as shown in \cref{fig:e2-sensitivity}. Its \auroc falls sharply under the attenuation, from $0.830$ at $\alpha=0$ to chance once the signal is fully replaced by background at $\alpha=1$, tracking the signal just as the distance and classical detectors do. All of these converge on $0.5$ as the discriminating structure is erased. The intermediate dip below $0.5$ near $\alpha=0.75$, shared with the Mahalanobis and isolation-forest scores, is itself a sign that the score reads the latent geometry, as a partly-morphed signal event lands deeper inside the background than a typical background event, and is ranked accordingly. The \qsvae's noise-insensitivity is therefore genuine selectivity. It responds to the physics that separates the signal from background and stops only when that physics is gone.

\section{Discussion}\label{sec:discussion}
The central result of our investigation is that the \qlif neuron sits off the accuracy--robustness frontier that the non-periodic classical spiking controls trace (\cref{fig:e4-capstone}). High-capacity detectors fit the clean background tightly and score accurately, and that same tight fit makes their scores swing when inputs are perturbed by noise. The bounded-classical clamp inherits this trade-off, becoming less robust as it is sharpened towards higher accuracy, while the \qlif neuron does not. We locate where the difference lives by measuring each model's input-to-score sensitivity directly. The \qsvae's anomaly scores respond $3.5$-fold less to a small geometry-respecting input perturbation than the least sensitive of the twelve classical-spiking controls'. Its score ordering, the quantity \auroc reads, is correspondingly stable where every control's score scrambles. This low sensitivity is consistent with, and quantitatively sufficient for, the retention gap. Within the classical band the local probe does not resolve the smaller retention differences, so we treat it as separating the \qsvae from the family rather than ordering the family itself. In the language of the robustness literature, the $\sin^2$ membrane acts as an architectural input-gradient regulariser where the low measured sensitivity is exactly the quantity that double backpropagation and Lipschitz-constrained training impose by an explicit penalty \citep{drucker1992improving,cisse2017parseval,hein2017formal}. It is obtained here from the neuron's dynamics with no extra loss term, and evaluated under a corruption-style protocol \citep{hendrycks2019benchmarking}.

The controls systematically eliminate the natural candidates for the property responsible for the \qsvae's robustness. Saturation is not enough to explain it, as the smooth-saturating control shares the vanishing derivative at both bounds, yet is as input-sensitive as the hard clamp. Gain contraction is not enough either. From \cref{eq:qlif} the \qlif per-step gain obeys $|\partial\alpha/\partial\theta|=\tfrac12|\sin(\theta+\mu)|\leq\tfrac12$ everywhere, and the trained layers respect it. This suggested a Lipschitz-style explanation, but the slope-matched logistic control realises the same gain profile and remains fully input-sensitive, showing that gain contraction alone cannot explain the observed robustness. What the eliminations leave is what the $\sin^2$ rotation alone possesses in the form of its periodic fold-back, which genuinely engages in the trained networks, together with the pure-state memory reconstruction that generates the map and places the leak inside the rotation as the memory angle $\mu$. Sine-based activations and Fourier-feature input encodings have been known to carry unusual inductive biases in classical networks \citep{ziyin2020neural,sitzmann2020implicit,tancik2020fourier}. The same $\sin^2$ structure makes data-reuploading quantum models Fourier series in their inputs \citep{schuld2021effect}, so the \qlif membrane sits in a well-documented family. The Periodic (external leak) control of \cref{sec:res-ablation} shows that the placement of the leak is not a necessary ingredient. With the leak moved outside the rotation, the neuron still reaches the \qsvae's robust corner. That control keeps the rest of the map, so whether the fold-back or the reconstruction it comes with carries the effect is not yet resolved. We acknowledge this as a limitation of this work, as a bounded periodic membrane causes the robustness, but a classical spiking network could adopt one directly. The quantum inspiration contributed the map itself, with its gain bound and its fold-back, found by taking the qubit dynamics seriously rather than by a search over activation functions. Because LHC anomaly detectors are trained on simulation and applied to data carrying extra detector noise, calibration drift, and mismodelling, robustness to input perturbation is a desirable property. Within the spiking pair the \fsvae's clean-data advantage over the \qsvae does not survive perturbation, so the quantum-inspired neuron is the more robust of the two. At a clean-accuracy cost the ablation shows to be the price of the dynamics rather than a tuning artefact.

This mechanism sits within a broader characterisation. We developed the \qsvae, the first fully spiking variational autoencoder built on the upgraded \qlif neuron, and benchmarked it against classical, spiking, and genuinely quantum detectors on a HEP collider task, with the published quantum baselines reproduced on the same data. On clean data simple methods win and complexity is not rewarded. Two methodological controls give insight to this, with the apparent robustness inversion under isotropic noise as a noise-model artefact and a signal-attenuation control confirming the robustness is genuine selectivity.

In addition, several limitations bound these findings. On clean data the classical-spiking \fsvae is the more accurate model, and the \qsvae is not a clean-accuracy state of the art competitor. Its value lies in the robustness regime, where searches care most about the low-false-positive figures of merit, and on clean data the \qsvae is among the weakest of the learned detectors there. Its compressed, high-floor code is the direct cause, and under the fair perturbation at $\sigma=0.3$ that ordering shifts within the spiking family. The \qsvae leads every probed spiking control on the low-false-positive metric, although the isolation forest still leads outright. No method wins on every signal, and on the most background-like signal a model-free statistic is competitive with all of them, which we also note as an interesting part of studies such as this.

The robustness that distinguishes the \qsvae is similar to regularised and compressed detectors like the isolation forest and quantum clustering methods, which are also robust. What is specific to the \qlif neuron is that it reaches this regime off the frontier that every non-periodic classical spiking neuron we tested obeys. One spiking alternative sits outside the control family and that is diversity in the per-neuron time constants which improves the robustness of classical spiking networks \citep{pereznieves2021neural}. That diversity cannot be the \qsvae's own mechanism, as every \qlif neuron shares one fixed leak. A classical spiking detector with learned heterogeneous leaks might nevertheless reach a comparably robust regime by that route, and testing one is a natural extension of the control family. If an isolation forest can be run, then it should be used. However, the result matters where the detector must spike, in the latency- and power-constrained trigger-adjacent settings for which neuromorphic hardware is built \citep{davies2018loihi,roy2019towards,dillon2026anomaly}.

We model the distribution shift with energy-matched, geometry-respecting Gaussian noise. This is a fair but simplified proxy where a real energy-scale system applies a coherent shift shared across an event and across the test population, which our per-event, per-coordinate draw does not capture. The comparison also spans only one dataset and one learned representation. We do not test whether an anomaly-score cut sculpts the di-jet invariant-mass spectrum. The standard trustworthiness check for a model-agnostic search, because the cached latents do not retain the constituent kinematics needed to reconstruct it. That check, with detector-level systematics, belongs to a dedicated deployment study. The genuinely quantum baselines are statevector-simulated on a single processor, which bounds the kernel to sixteen qubits and omits it at the largest latent dimension.

We are explicit about what is and is not quantum throughout this work. The \qlif neuron is quantum-inspired, since its membrane follows single-qubit population dynamics in a pure-state ansatz and it is a hardware-feasible spiking nonlinearity, while every operation we run is classically simulable and we claim no computational speed-up. The kernel and autoencoder baselines are genuine quantum machine-learning models, compared against each other on equal data and compute. Within that frame the reproduced quantum kernel beats its own classical reference, as originally reported, but not the strongest classical or spiking detectors. Its advantage also depends on the latent dimension. The contribution is a fair placement of these methods, and a mechanism-pinned characterisation of the quantum-inspired neuron, rather than a demonstration of any quantum advantage.

Three directions follow from this work. The \qlif neuron's update is a scalar recurrence that a programmable neuromorphic substrate could plausibly implement, though it is not a native LIF primitive. A natural next step would be a hardware demonstration measuring the \qsvae's energy and latency against the classical-spiking baseline. The robustness result lays a foundation for tests under realistic, correlated detector shift in place of the Gaussian proxy, and across a second dataset and representation to establish how far the capacity-decoupled regime can generalise. Because the $\sin^2$ dynamics create robustness at the cost of capacity, a neuron that relaxes its saturation where capacity is needed while retaining it where robustness is needed would be worth pursuing as an ideal form. The result we carry forward is that a cheap, classically simulable, quantum-inspired spiking neuron occupies a distinctive and useful corner of the accuracy--robustness plane. It is robust under our proxy for the noisy, shifted conditions of deployment, that the non-periodic classical spiking neurons we tested do not reach and should be further explored.

\section{Methods}\label{sec:methods}

\subsection{Data and task}
\label{sec:meth-data}
We use the di-jet latent representation released with the quantum anomaly-detection study of Belis \textit{et al.}~\citep{belis2024quantum}. In that work a classical convolutional autoencoder compresses each LHC di-jet event (two jets, each described by up to one hundred constituents with features $\Delta\eta$, $\Delta\phi$, and $p_T$) into a per-jet latent vector. We concatenate the two jets into a single event vector of dimension $2\lat$, where $\lat\in\{4,8,16,32\}$ is the per-jet latent dimension. The autoencoder uses a $\tanh$ bottleneck, so latent coordinates lie in the range $(-1,1)$. The background is the Standard-Model QCD multijet production. The three signals beyond the Standard Model are a narrow ($3.5\,\mathrm{TeV}$) and a broad ($1.5\,\mathrm{TeV}$) Randall--Sundrum graviton decaying to $WW$, and a scalar $A\!\to\!HZ\!\to\!ZZZ$ at $3.5\,\mathrm{TeV}$. Each detector is trained as one-class models on a withheld sample of background events and evaluated on a balanced mixture of background and each signal in turn. We report per-signal scores and an Overall score pooled across the three signals. The spiking models draw from a $15$k-event background pool, of which $15\%$ is held out for validation. The classical detectors fit about $25$k events, and the quantum-method budgets are given in \cref{tab:landscape}.

\subsection{Anomaly-detection protocol and metrics}
\label{sec:meth-metrics}
The primary metric is the area under the ROC curve (\auroc). To probe the low-false-positive regime that matters for a search we additionally report the inverse background efficiency $1/\varepsilon_b$ at fixed signal efficiency $\varepsilon_s\in\{0.6,0.8\}$, following Belis \textit{et al.}~\citep{belis2024quantum}, and the significance improvement characteristic $\varepsilon_s/\sqrt{\varepsilon_b}$ at fixed background efficiency $\varepsilon_b\in\{10^{-2},10^{-3}\}$, following the Dark Machines challenge~\citep{aarrestad2022darkmachines}. These are reported for every detector with saved per-event scores in \cref{tab:supp-search}, with per-signal \auroc in \cref{tab:supp-persignal-l16}. Where repeated fits exist, we report the mean and across-seed standard deviation. The headline spiking detectors use five seeds, while the ablation sweeps use three or five as stated in \cref{sec:res-ablation}, and the quantum detectors use one or five depending on dimension, and the clean hybrid, PQAE, and BQAE results use three. The PQAE/BQAE robustness curves are fixed-seed stress checks. The single-fit model-free and classical detectors have no seed variation, so for them we instead report a bootstrap $95\%$ confidence interval over the test set. For the quantum kernel and quantum autoencoders, whose cost limits the training-set size, each seed draws a background-only training subset from a fixed pool.

\subsection{The spiking autoencoder and the ablation neurons}\label{sec:meth-qlif}
The \qlif neuron and the quantum-inspired autoencoder are derived in \cref{sec:model-qlif,sec:model-qsvae}. Here we provide the relevant reproduction details and the ablation neurons. Both spiking detectors use our feature-vector adaptation of the fully spiking variational autoencoder of Kamata \textit{et al.}~\citep{kamata2022fsvae}. The model consumes the $2\lat$-dimensional latent vector directly, as a spiking multilayer-perceptron encoder and decoder run over $T=16$ time steps, with a Bernoulli spiking posterior and prior. It follows the evidence-lower-bound framework of variational autoencoders \citep{kingma2014vae}, with the latent Kullback--Leibler term replaced by the \fsvae-specific maximum-mean-discrepancy term between samples from the posterior and prior spike trains \citep{kamata2022fsvae}, which is used as the per-event anomaly score. The two detectors differ only in the neuron, where the \fsvae uses the standard LIF unit and the \qsvae the \qlif neuron of \cref{eq:qlif}. Both networks train by backpropagation through time with surrogate gradients~\citep{neftci2019surrogate} and threshold-dependent batch normalisation~\citep{zheng2021tdbn}. For the \qlif neuron we clamp the backward membrane gradient (\cref{eq:qlif-grad}) for stable optimisation and initialise the threshold-dependent normalisation to a firing rate near $12\%$. After training the mean firing rates settle higher and differ by neuron. In the three-seed probe we have \fsvae at $0.29$, \qsvae at $0.23$, the dense Clamped neuron at $0.37$, and the firing-matched Clamped~(matched) at $0.19$. It is these trained rates that the firing-matched control approximates. The \qlif spike threshold is $0.75$ on the excited-state population and its leak is set by $T_1=60$ steps, as in the original \qlif~\citep{brand2024qlif}. The architecture and full hyperparameter set are detailed in \cref{tab:supp-hyper}. Both detectors train for $60$ epochs with the AdamW optimiser, learning rate of $10^{-3}$, weight decay of $10^{-3}$, and batch size of $256$, converging before evaluation.

To isolate which property of the \qlif neuron produces the robustness (\cref{sec:res-ablation}), we introduce a bounded-classical control, the Clamped neuron. It keeps the \qlif neuron's bounded $[0,1]$ membrane, spike threshold, and leak, but replaces the $\sin^2$ rotation of \cref{eq:qlif} with ordinary linear leaky integration followed by a hard clamp, $\alpha[t+1]=\mathrm{clip}(\lambda\,\alpha[t]+\theta[t],\,0,\,1)$, where $\theta[t]$ is the weighted input and $\lambda$ the same leak factor. It is bounded exactly as the \qlif neuron is, but its dynamics are classical and locally linear, so comparing it to the \qsvae separates the bounded membrane from the $\sin^2$ dynamics. Because a sparser neuron is mechanically more robust, we sweep the clamp's threshold-dependent normalisation scale ($\texttt{bn\_scale}\in\{0.30,0.45,0.60,0.80,1.0\}$), which sets the trained firing rate as $0.37$ at the densest setting, $0.19$ at $\texttt{bn\_scale}=0.45$, and sparser still below that. This $\texttt{bn\_scale}=0.45$ setting, Clamped (matched), fires close to the \qsvae's rate ($0.19$ against $0.23$) and removes firing density as a confound. To test whether smooth saturation alone reproduces the effect, the smooth-saturating neuron replaces the hard clamp with a logistic map $\alpha[t+1]=\sigma\!\big(g(\lambda\,\alpha[t]+\theta[t]-\tfrac12)\big)$, bounded to $(0,1)$ with a derivative that vanishes at both bounds like the $\sin^2$ map but monotone and non-periodic. We train it at three settings: gain $g=4$ (central slope $g/4=1$, matching the clamp's interior gain), gain $g=2$ (maximum slope one half, matching the \qlif map's global gain bound $|\partial\alpha/\partial\theta|\leq\tfrac12$), and $g=4$ with the normalisation lowered to $\texttt{bn\_scale}=0.45$ (firing-matched, mirroring the matched clamp). And to test ordinary regularisation, we train the \fsvae at weight decay $\{10^{-2},3\times10^{-2},10^{-1}\}$, up to $100$ times the $10^{-3}$ default. All controls share the \qlif neuron's threshold, leak, optimiser, and schedule. Their backward rules are not identical, as the \qlif and periodic neurons share a capped recurrent gradient that passes through the hard reset. The clamped and smooth-saturating controls zero the recurrent gradient at a reset through ordinary automatic differentiation, and the \fsvae keeps its published surrogate and multiplicative reset. The firing, saturation, and regularisation sweeps use three seeds and the other variants five.

\subsection{Baseline families}
\label{sec:meth-baselines}
We compare against four families of baselines, all consuming the same latents. Model-free statistics are computed directly on the latent vector with no training, with metrics including the latent energy ($\mathrm{mean}\,z^2$), the $L_2$ norm, and the maximum absolute coordinate \maxz. Classical detectors are taken from the PyOD library~\citep{zhao2019pyod}, namely a $k$-nearest-neighbour distance, a Mahalanobis distance to the background, the isolation forest~\citep{liu2008iforest}, a deep autoencoder, and the deep one-class detector Deep SVDD~\citep{ruff2018deepsvdd}. Quantum detectors are the suite of Belis \textit{et al.}~\citep{belis2024quantum}, reimplemented in PennyLane~\citep{bergholm2018pennylane}. The quantum kernel is built from a data-reuploading feature map, in which each qubit encodes two latent coordinates through rotations, with nearest-neighbour CNOT entanglement and $L$ reuploading repetitions. We evaluate the kernel as the squared state overlap $k(x,x')=|\langle\psi(x)|\psi(x')\rangle|^2$ from statevectors, and pass it to a one-class support vector machine with a precomputed Gram matrix~\citep{havlicek2019supervised,schuld2019quantum}. The family also includes a non-entangled variant of that kernel, the quantum $k$-means and $k$-medians clustering scores, and the classical RBF one-class support vector machine that the original study uses as its reference. The Gram matrix scales as $N^2$, which sets the training budget to about $600$ events. Quantum autoencoders comprise two fully quantum variational autoencoders, a persistent (PQAE) and a bottleneck (BQAE) variant scored by reconstruction error, and a hybrid autoencoder pairing a quantum data-reuploading encoder with a classical decoder, in entangled and non-entangled forms. The fully quantum variants are statevector-feasible only at the smaller widths (the PQAE through $\lat=8$ and the BQAE at $\lat=4$). We validated the quantum suite against the per-signal numbers and the entanglement and depth scaling reported by Belis \textit{et al.} (\cref{fig:quantum-reps}).

\subsection{Perturbation protocols: noise robustness and sensitivity}\label{sec:meth-noise}
To probe robustness to noise and distribution shift we fit every detector on the clean background and, at evaluation only, added noise to each test latent before scoring. Identical noised inputs are scored by every method, so differences reflect the detectors rather than the perturbation. We use two noise models at matched magnitude. Isotropic noise adds independent Gaussian noise of standard deviation $\sigma$ to every coordinate equally. Geometry-respecting noise instead scales the noise per coordinate by the background's own spread in that coordinate, $\sigma_i=\sigma\,s_i$ with $s_i=\hat\sigma_i^{\mathrm{bkg}}/\sqrt{\langle(\hat\sigma^{\mathrm{bkg}})^2\rangle}$, so the scale factors are root-mean-square normalised ($\langle s_i^2\rangle=1$). The total injected power equals the isotropic case at the same $\sigma$, only redistributed to follow the data, which makes the two models a fair comparison rather than one being merely smaller. We sweep $\sigma$ over $\{0,0.05,0.1,0.15,0.2,0.3\}$ and report the \auroc at each level and the retention $\mathrm{AUROC}(\sigma)/\mathrm{AUROC}(0)$. The geometry-respecting model is the fair baseline, and the isotropic model is retained to expose the noise-model artefact of \cref{sec:res-fairnoise}. The robustness and sensitivity pipelines re-score every detector on their own evaluation split, so the clean \auroc values quoted in the robustness sections differ from \cref{tab:landscape} by at most $0.010$. Retention always uses the pipeline's own clean value as its denominator.

A complementary sensitivity control tests whether a detector's noise-robustness is genuine signal-tracking or just insensitivity. Leaving the trained model fixed, we attenuate each test-signal event towards the background by linearly interpolating it with a randomly paired background event, $x\mapsto(1-\alpha)\,x+\alpha\,x_{\mathrm{bkg}}$, sweeping $\alpha$ over $\{0,0.25,0.5,0.75,1\}$ from the true signal ($\alpha=0$) to a signal statistically identical to background ($\alpha=1$), while background test events are left unchanged. A detector that reads the discriminating physics must lose separation towards $\mathrm{AUROC}=0.5$ as $\alpha\to1$, whereas one insensitive to its input would not. As a further degeneracy check we record the spread of the min--max-normalised anomaly score at $\alpha=0$, which collapses to zero for a near-constant detector.

To probe the robustness mechanism directly, we measure how strongly each trained model's anomaly score $S$ responds to a small input perturbation. Holding the model fixed, we perturb every test latent by a geometry-respecting step $\delta=\sigma\,s\odot\eta$ (with $\eta\sim\mathcal{N}(0,I)$ and the same per-coordinate scale $s$ as above, at probe scales $\sigma\in\{0.05,0.1,0.2\}$). From this we record the mean absolute score change $\langle|S(x+\delta)-S(x)|\rangle$, averaged over draws and test events and normalised by the model's own clean score spread $\operatorname{std}(S(x))$. This normalised sensitivity of the score to the input is comparable across models whose scores have different scales, and a smaller value means the anomaly scores are more decoupled from input perturbation. Because the normalisation could itself differ across models, we corroborate it with two normalisation-free readouts of the same perturbation: the Kendall rank correlation between the clean and perturbed score orderings, and the mean absolute rank displacement as a fraction of the test set. The ordering, rather than the score values, is what \auroc depends on. We report these, together with the trained mean spiking-layer firing rate averaged over three seeds, for the \qsvae and each of the twelve non-periodic classical-spiking controls, all computed from the saved checkpoints. Finally, to characterise the operating regime of the trained neurons we replay the recorded spike-layer inputs through each membrane recurrence and histogram the per-step drive and its local gain. For this, the \qlif drive angle is $\theta+\mu$ and the gain is $\tfrac12|\sin(\theta+\mu)|$, for the smooth-saturating control the gain is of its logistic map, and for the clamp the fraction of updates is inside its linear interior.

\subsection{Compute}
\label{sec:meth-compute}
All experiments run on a single NVIDIA RTX 3080 Ti. The quantum detectors are simulated by statevector with PennyLane~\citep{bergholm2018pennylane}. The quantum kernel is feasible through $\lat=16$ (sixteen qubits) and is omitted at $\lat=32$, where the $2^{32}$-dimensional statevector is intractable, leaving the clustering and reference detectors at that width. The fully quantum autoencoders are simulated at their statevector-feasible widths: eight and sixteen qubits for the PQAE at $\lat=4,8$, and fifteen qubits for the BQAE at $\lat=4$. Code, configurations, curated derived results, saved spiking-model checkpoints and event scores, and quantum-autoencoder parameters and event scores are released with the paper (see Code availability).

\begin{figure}[!htb]
  \centering
  \includegraphics[width=0.92\linewidth]{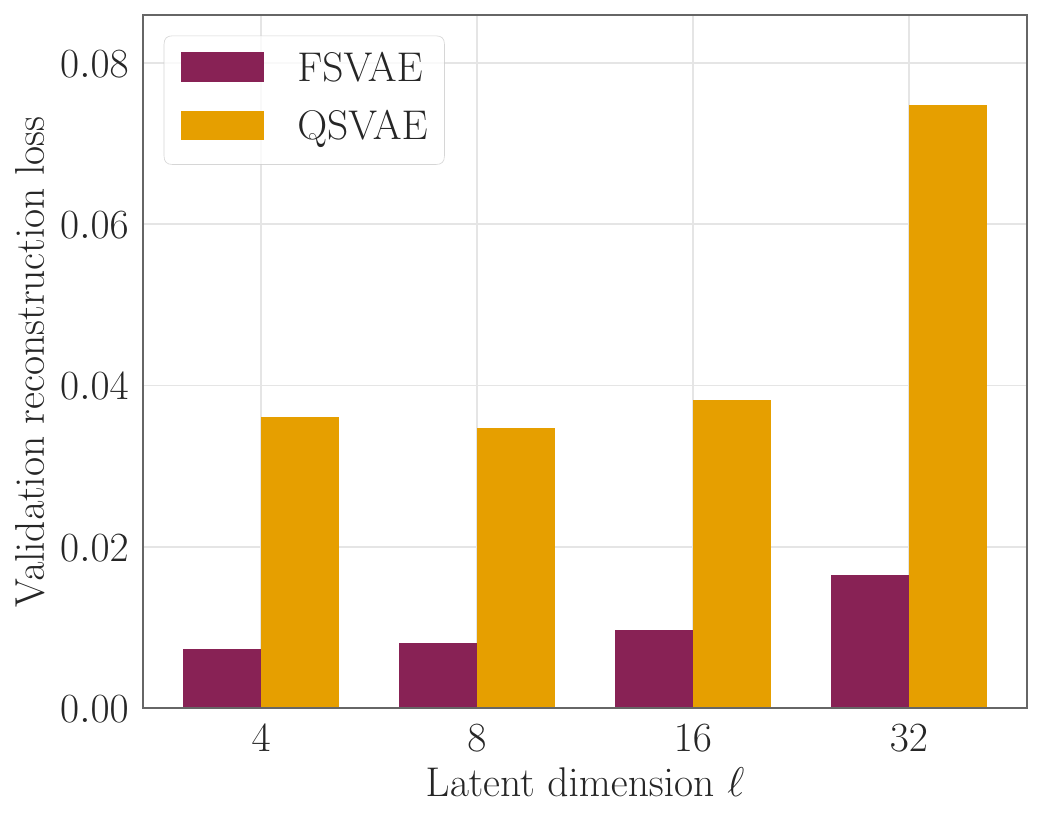}
  \caption{\textbf{Reconstruction floor.} Final validation reconstruction loss of the \fsvae and \qsvae across latent dimension. The \qlif neuron's bounded $\sin^2$-rotation membrane settles at a ${\sim}4$ times higher reconstruction floor than the classical-leaky neuron. The firing-matched ablation (\cref{sec:res-ablation}) shows this floor carries no accuracy penalty, since an equally-bounded classical neuron reconstructs as well as the \fsvae. Instead it is consistent with a dynamics-specific regularisation that we associate with the \qsvae's robustness and stability.}
  \label{fig:reconfloor}
\end{figure}


\section*{Data availability}\label{sec:data}
The di-jet latent-space data are from Belis \textit{et al.}~\citep{belis2024quantum} and are publicly available at Zenodo, \href{https://doi.org/10.5281/zenodo.7673769}{10.5281/zenodo.7673769}.

\section*{Code availability}\label{sec:code}
The anomaly-detection pipeline, the \qlif neuron and \qsvae implementation, and the classical, quantum, and quantum-autoencoder baseline reproductions are available at \url{https://github.com/deanbrand/QSVAE}. Version~1.0.1 of the software, together with experiment configurations, curated derived results, saved spiking-model checkpoints and event scores, and quantum-autoencoder parameters and event scores, is archived at Zenodo, \href{https://doi.org/10.5281/zenodo.24460595}{\nolinkurl{10.5281/zenodo.24460595}}. Other baseline event scores can be regenerated using the supplied experiment drivers and the public input dataset.

\section*{Acknowledgements}\label{sec:ack}
This work was supported by the National Institute for Theoretical and Computational Sciences. C.M.M acknowledges the support from the African Institute for Mathematical Sciences Research and Innovation Centre of Rwanda.

\section*{Author contributions}\label{sec:contrib}
D.B. conceived the study, developed the \qlif neuron and the software, ran the experiments, performed the formal analysis, and wrote the original draft. D.S. contributed the persistent and bottleneck quantum-autoencoder baselines, and C.M.M. contributed the hybrid quantum-autoencoder baseline. F.P. supervised the work and acquired funding. All authors reviewed and edited the manuscript.

\section*{AI Declaration}\label{sec:AI_declaration}
The authors used OpenAI’s GPT-6 Astra and Anthropic’s Claude Fable 5.1 to assist with manuscript preparation and the development of computational code. These tools were used in accordance with the authors’ research decisions, directions, and intended methodology. The tools also assisted with language editing and code review. All AI-assisted outputs and generated code were critically reviewed by the authors to ensure their correctness and alignment with the intended research objectives and methodology. The authors take full responsibility for the final manuscript and all reported results.

\section*{Competing interests}\label{sec:coi}
The authors declare no competing interests.

\clearpage
\appendix
\renewcommand{\thefigure}{S\arabic{figure}}\renewcommand{\theHfigure}{S\arabic{figure}}\setcounter{figure}{0}
\renewcommand{\thetable}{S\arabic{table}}\renewcommand{\theHtable}{S\arabic{table}}\setcounter{table}{0}
\section{Supplementary material}\label{sec:supp}

\subsection{Architecture and training hyperparameters}\label{sec:supp-hyper}
The spiking variational-autoencoder configuration is shared by the \fsvae and the \qsvae; the two differ only in the spiking neuron (\cref{tab:supp-hyper}). Values are the run configuration in \texttt{config.py} (the per-jet latent width is $2\lat$ for $\lat\in\{4,8,16,32\}$).

\begin{table*}[h]
  \centering
  \caption{\textbf{Spiking-VAE architecture and training settings} (FSVAE and QSVAE), with the \qlif neuron parameters. The FSVAE uses a standard leaky integrate-and-fire neuron in place of \qlif.}
  \label{tab:supp-hyper}
  \begin{tabular}{l l}
    \toprule
    Component & Setting \\
    \midrule
    Input width (di-jet latent)      & $2\lat$ \\
    Spiking-VAE latent dimension     & $64$ \\
    Encoder/decoder hidden widths    & $256, 128$ \\
    Time steps $T$                   & $16$ \\
    Spiking neuron                   & LIF (FSVAE) / \qlif (QSVAE) \\
    Objective                        & ELBO (reconstruction $+$ MMD) \\
    Optimiser                        & AdamW, learning rate $10^{-3}$, weight decay $10^{-3}$ \\
    Batch size                       & $256$ \\
    Epochs                           & $60$ \\
    Seeds                            & $5$ spiking \\
    \midrule
    \multicolumn{2}{l}{\textit{QLIF neuron}} \\
    Firing threshold                 & $0.75$ \\
    $T_1$ relaxation                 & $60$ \\
    Membrane time constant $\tau$    & $1$ \\
    tdBN angle scale (firing ${\approx}12\%$) & $1.0$ \\
    Graded (non-binary) output       & no \\
    Backward membrane-gradient clamp & $1.0$ \\
    \bottomrule
  \end{tabular}
\end{table*}

\subsection{Reproduction validation of the quantum baselines}\label{sec:supp-repro}
\Cref{tab:repro-quantum} places our PennyLane reimplementation of the quantum kernel and its classical reference next to the published per-signal values of Belis \textit{et al.}~\citep{belis2024quantum}, read from the printed AUC legends of their Fig.~3. Every per-signal \auroc agrees to within $0.065$, the largest gap on the near-chance broad graviton, where both implementations bracket $0.5$. Belis \textit{et al.} report no signal-pooled Overall \auroc, so the Overall figures quoted in the main text are our own aggregate rather than a reproduced published value; our classical RBF reference is somewhat weaker than theirs, so our measured quantum-over-classical margin is, if anything, conservative.

\begin{table*}[!htb]
  \centering
  \caption{\textbf{Reproduction validation of the quantum kernel baseline against Belis \textit{et al.}} Per-signal \auroc of the unsupervised quantum kernel machine (nearest-neighbour entanglement) and its classical RBF one-class support vector machine reference, at the seed paper's reference operating point ($N_\mathrm{train}=600$, $\nu=0.01$; $\lat$ equals the qubit count). Published values are read from the printed AUC legends of Fig.~3 of Belis \textit{et al.}~\citep{belis2024quantum}: their per-signal panel fixes $\lat=8$, and their latent-dimension panel reports the $A\!\to\!HZ$ signal, which we match at $\lat=16$. Our values are five-seed means from \texttt{runs/hep\_quantum\_l8} and
  \texttt{runs/hep\_quantum\_l16}.}
  \label{tab:repro-quantum}
  \sisetup{table-format=1.3}
  \begin{tabular}{l l S S S}
    \toprule
    Detector & Signal ($\lat$) & {Belis \textit{et al.}} & {This work} & {$|\Delta|$} \\
             &                 & {(published)}           & {(reproduced)} & \\
    \midrule
    \multicolumn{5}{l}{\textit{Unsupervised quantum kernel machine (entangled)}} \\
    & Narrow $G\!\to\!WW$ ($\lat{=}8$)  & 0.995 & 0.997 & 0.002 \\
    & $A\!\to\!HZ$ ($\lat{=}8$)         & 0.947 & 0.968 & 0.021 \\
    & Broad $G\!\to\!WW$ ($\lat{=}8$)   & 0.476 & 0.541 & 0.065 \\
    & $A\!\to\!HZ$ ($\lat{=}16$)        & 0.968 & 0.980 & 0.012 \\
    \addlinespace
    \multicolumn{5}{l}{\textit{Classical RBF one-class support vector machine (the kernel's own reference)}} \\
    & Narrow $G\!\to\!WW$ ($\lat{=}8$)  & 0.993 & 0.986 & 0.007 \\
    & $A\!\to\!HZ$ ($\lat{=}8$)         & 0.933 & 0.889 & 0.044 \\
    & Broad $G\!\to\!WW$ ($\lat{=}8$)   & 0.456 & 0.441 & 0.015 \\
    & $A\!\to\!HZ$ ($\lat{=}16$)        & 0.912 & 0.848 & 0.064 \\
    \bottomrule
  \end{tabular}
\end{table*}

\subsection{Full benchmark tables}\label{sec:supp-tables}
\Cref{tab:supp-overall} gives the Overall \auroc for every detector across all four latent dimensions, including the matched-scale ($N_\mathrm{train}\!\approx\!600$) spiking rows and the persistent and bottleneck quantum autoencoders at their statevector-feasible widths. \Cref{tab:supp-persignal-l16} gives the per-signal \auroc and the search-oriented low-false-positive metric $1/\varepsilon_b$ at $\varepsilon_s=0.8$ at $\lat=16$.

\subsection{Fully quantum autoencoders}\label{sec:supp-qae}
With the angle-domain reconstruction described in \cref{sec:meth-baselines}, the three-seed clean Overall \auroc is $0.836\pm0.005$ for PQAE and $0.839\pm0.004$ for BQAE at $\lat=4$, and $0.853\pm0.005$ for PQAE at $\lat=8$. The broad graviton is hardest in all three configurations ($0.580\pm0.006$, $0.573\pm0.009$, and $0.584\pm0.014$, respectively). In the fixed-seed robustness check, the fraction of clean Overall \auroc retained at $\sigma=0.3$ is $0.812$ for PQAE and $0.767$ for BQAE at $\lat=4$, and $0.814$ for PQAE at $\lat=8$. No clean coordinate crosses the $[-1,1]$ encoder range. At $\sigma=0.3$, the pre-clipping coordinate rates at $\lat=4$ are $3.95\%$ for background and $5.33\%$, $2.91\%$, and $3.42\%$ for narrow $G$, broad $G$, and $A\!\to\!HZ$; at $\lat=8$ they are $6.81\%$, $10.27\%$, $6.79\%$, and $9.23\%$. We clip these values at the boundary and never wrap them periodically. The clean and noised results come from \texttt{runs/hep\_qae\_v2\_l\{4,8\}} and \texttt{runs/hep\_robustness\_qae\_v2}.

\begin{figure*}[!htb]
  \centering
  \includegraphics[width=0.49\linewidth]{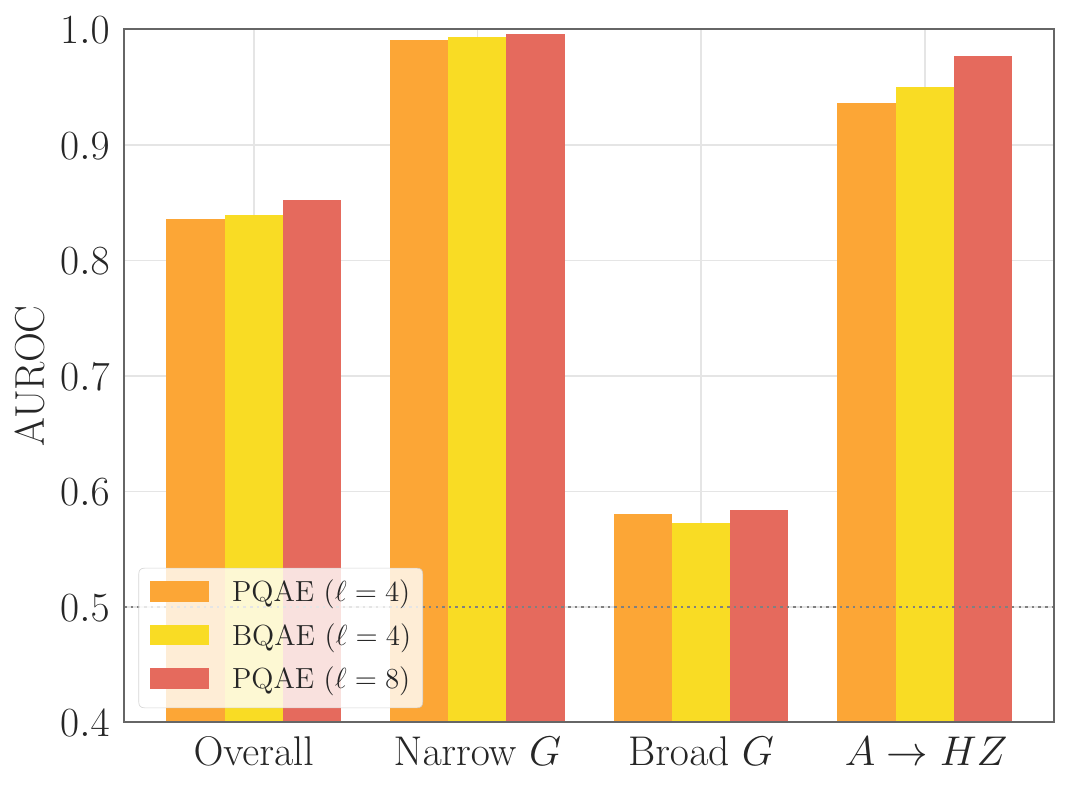}\hfill
  \includegraphics[width=0.49\linewidth]{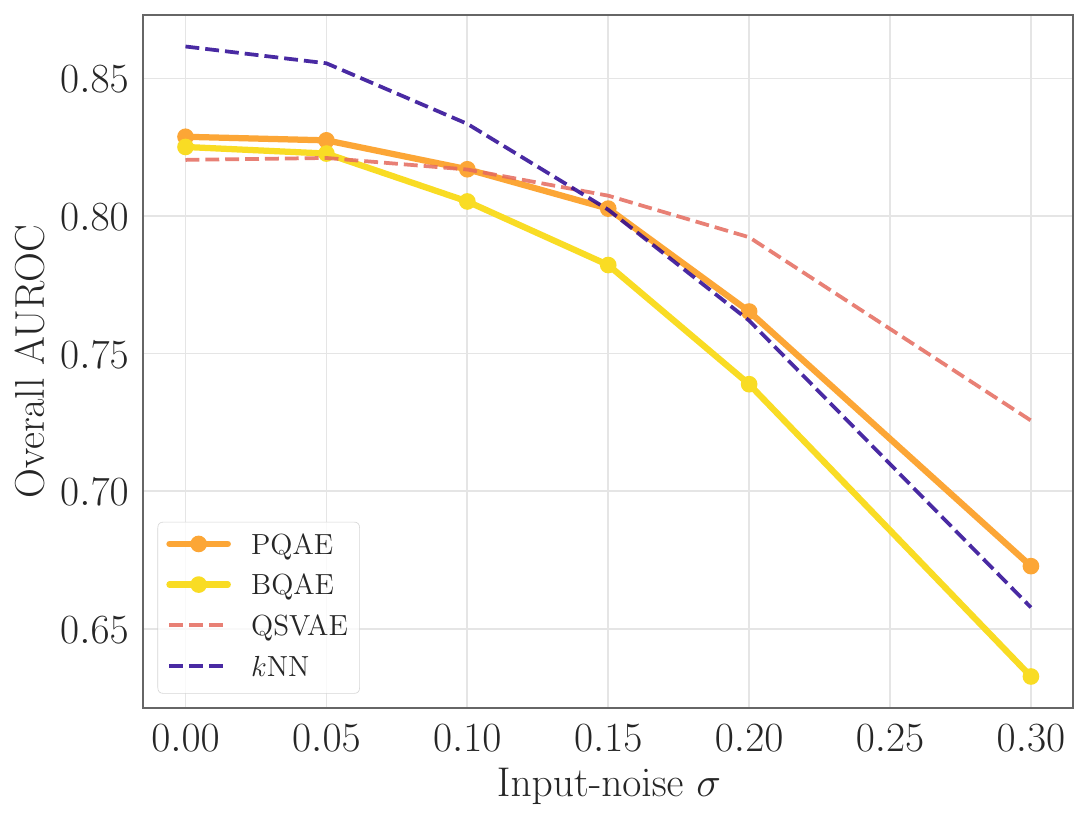}
  \caption{\textbf{Fully quantum autoencoders.} \textbf{(a)}~Clean per-signal \auroc for PQAE and BQAE at their statevector-feasible widths (three-seed means). Broad $G$ remains the hardest signal, and PQAE improves from $\lat=4$ to $\lat=8$. \textbf{(b)}~At $\lat=4$, input noise reduces the single-seed PQAE and BQAE scores more sharply than the regularised \qsvae control; $k$NN is included as a clean-data leader. Every model is trained on clean QCD background and then scored on the same noised test events.}
  \label{fig:supp-qae}
\end{figure*}


\begin{table*}[!htb]
  \centering
  \caption{\textbf{Overall \auroc across latent dimension.} Compact uncertainty notation: $0.877(5)$ denotes $0.877\pm0.005$, the across-seed standard deviation in units of the last digit (five seeds for the spiking models; one or five for the quantum methods, depending on dimension; and three for the quantum autoencoders); no parenthesis means a single fit. The matched-scale rows FSVAE\,(600) and QSVAE\,(600) repeat the spiking models at the quantum kernel's $\sim\!600$-event budget. ``--'' marks a configuration that is statevector-infeasible at that dimension. Source: \texttt{runs/hep\_comparison\_v2\_l\{4,8\}} and \texttt{runs/hep\_comparison\_l\{16,32\}}.}
  \label{tab:supp-overall}
  \begin{tabular}{l l c c c c }
    \toprule
    Family & Method & $\ell=4$ & $\ell=8$ & $\ell=16$ & $\ell=32$ \\
    \midrule
    model-free & $\max|z|$ & 0.468 & 0.866 & 0.719 & 0.500 \\
     & $L_2$ energy & 0.516 & 0.721 & 0.791 & 0.833 \\
    \midrule
    classical & $k$NN & 0.867 & 0.892 & 0.891 & 0.881 \\
     & Mahalanobis & 0.833 & 0.869 & 0.879 & 0.865 \\
     & Isolation Forest & 0.835 & 0.849 & 0.846 & 0.855 \\
     & AutoEncoder & 0.825 & 0.855 & 0.874 & 0.868 \\
     & Deep SVDD & 0.851 & 0.770 & 0.684 & 0.799 \\
    \midrule
    spiking & FSVAE & 0.824(5) & 0.865(2) & 0.877(5) & 0.864(8) \\
     & QSVAE & 0.821(2) & 0.840(2) & 0.827(1) & 0.813(3) \\
     & FSVAE (600) & -- & 0.841(9) & 0.842(3) & -- \\
     & QSVAE (600) & -- & 0.837(3) & 0.821(2) & -- \\
    \midrule
    quantum & Q-kernel (Ent.) & 0.779 & 0.835(2) & 0.855(3) & -- \\
     & Q-kernel (no ent.) & 0.718 & 0.824(5) & 0.829(4) & -- \\
     & QK-medians & 0.852 & 0.863(3) & 0.855(2) & 0.803 \\
     & QK-means & 0.835 & 0.859(1) & 0.848(1) & 0.815 \\
     & RBF-OCSVM (ref.) & 0.762 & 0.772(30) & 0.755(34) & 0.838 \\
    \midrule
    quantum-AE & Hybrid QAE (non-ent.) & 0.816(6) & 0.862(8) & 0.877(8) & 0.844(9) \\
     & Hybrid QAE (Ent.) & 0.832(6) & 0.873(3) & 0.866(5) & 0.816(1) \\
     & PQAE & 0.836(5) & 0.853(5) & -- & -- \\
     & BQAE & 0.839(4) & -- & -- & -- \\
    \bottomrule
  \end{tabular}
\end{table*}

\begin{table*}[!htb]
  \centering
  \caption{\textbf{Per-signal performance at $\ell=16$.} \auroc ($\pm$ across-seed std where multi-seed) for each signal, and the search-oriented inverse background efficiency $1/\varepsilon_b$ at signal efficiency $\varepsilon_s=0.8$ (Overall). Narrow and broad denote the narrow and broad Randall--Sundrum gravitons; $A\!\to\!HZ$ the scalar. Source: \texttt{runs/hep\_comparison\_l16}.}
  \label{tab:supp-persignal-l16}
  \begin{tabular}{l l c c c c}
    \toprule
    Family & Method & Narrow & Broad & $A\!\to\!HZ$ & $1/\varepsilon_b@0.8$ \\
    \midrule
    model-free & $\max|z|$ & 0.683 & 0.667 & 0.807 & 1.9 \\
     & $L_2$ energy & 0.973 & 0.520 & 0.879 & 2.2 \\
    \midrule
    classical & $k$NN & 0.998 & 0.689 & 0.987 & 5.6 \\
     & Mahalanobis & 0.996 & 0.663 & 0.978 & 4.9 \\
     & Isolation Forest & 0.996 & 0.573 & 0.968 & 3.2 \\
     & AutoEncoder & 0.995 & 0.649 & 0.977 & 4.7 \\
     & Deep SVDD & 0.865 & 0.441 & 0.746 & 1.8 \\
    \midrule
    spiking & FSVAE & 0.997$\pm$0.001 & 0.657$\pm$0.009 & 0.977$\pm$0.006 & 4.4 \\
     & QSVAE & 0.991$\pm$0.000 & 0.544$\pm$0.003 & 0.945$\pm$0.001 & 2.8 \\
     & FSVAE (600) & 0.991$\pm$0.001 & 0.603$\pm$0.011 & 0.933$\pm$0.003 & 3.1 \\
     & QSVAE (600) & 0.990$\pm$0.000 & 0.535$\pm$0.004 & 0.938$\pm$0.002 & 2.6 \\
    \midrule
    quantum & Q-kernel (Ent.) & 0.999$\pm$0.001 & 0.587$\pm$0.008 & 0.980$\pm$0.003 & 3.7 \\
     & Q-kernel (no ent.) & 0.989$\pm$0.002 & 0.549$\pm$0.004 & 0.949$\pm$0.006 & 2.8 \\
     & QK-medians & 0.993$\pm$0.001 & 0.617$\pm$0.008 & 0.956$\pm$0.004 & 3.7 \\
     & QK-means & 0.994$\pm$0.001 & 0.598$\pm$0.001 & 0.953$\pm$0.003 & 3.4 \\
     & RBF-OCSVM (ref.) & 0.978$\pm$0.012 & 0.440$\pm$0.035 & 0.848$\pm$0.054 & 1.8 \\
    \midrule
    quantum-AE & Hybrid QAE (non-ent.) & 0.997$\pm$0.001 & 0.652$\pm$0.015 & 0.983$\pm$0.007 & 4.4 \\
     & Hybrid QAE (Ent.) & 0.996$\pm$0.001 & 0.630$\pm$0.013 & 0.973$\pm$0.004 & 4.2 \\
     & PQAE & -- & -- & -- & -- \\
     & BQAE & -- & -- & -- & -- \\
    \bottomrule
  \end{tabular}
\end{table*}

\begin{table*}[!htb]
  \centering
  \caption{\textbf{Uncertainty and low-false-positive search metrics ($\ell=16$, Overall).} \auroc with, for the single-fit detectors, a bootstrap $95\%$ confidence interval over the test set ($1000$ resamples); for the spiking detectors, the across-seed standard deviation (five seeds). $1/\varepsilon_b$ is the inverse background efficiency at signal efficiency $\varepsilon_s$; $\mathrm{SIC}=\varepsilon_s/\sqrt{\varepsilon_b}$ is the significance improvement at background efficiency $\varepsilon_b$. Computed from the saved per-event test scores. Source: \texttt{runs/hep\_score\_stats}.}
  \label{tab:supp-search}
  \begin{tabular}{l l c c c c}
    \toprule
    Method & \auroc & $1/\varepsilon_b@0.6$ & $1/\varepsilon_b@0.8$ & $\mathrm{SIC}@10^{-2}$ & $\mathrm{SIC}@10^{-3}$ \\
    \midrule
    FSVAE & $0.877\pm0.005$ & 60.1 & 4.4 & 5.5 & 9.9 \\
    QSVAE & $0.827\pm0.001$ & 19.2 & 2.8 & 4.0 & 5.7 \\
    Clamped & $0.876\pm0.005$ & 63.9 & 4.4 & 5.5 & 9.5 \\
    Clamped (matched) & $0.883\pm0.003$ & 54.7 & 4.9 & 5.4 & 9.7 \\
    Isolation Forest & $0.847\,[0.844,0.850]$ & 30.9 & 3.2 & 4.9 & 9.0 \\
    Mahalanobis & $0.879\,[0.876,0.882]$ & 51.7 & 5.0 & 5.3 & 2.0 \\
    $L_2$ energy & $0.791\,[0.787,0.795]$ & 9.7 & 2.2 & 3.1 & 3.8 \\
    $\max|z|$ & $0.719\,[0.715,0.723]$ & 3.8 & 1.9 & 1.1 & 0.9 \\
    \bottomrule
  \end{tabular}
\end{table*}

\begin{table*}[!htb]
  \centering
  \caption{\textbf{Classical-spiking ablation controls ($\ell=16$).} Clean \auroc from the robustness pipeline (its own evaluation split; values differ from \cref{tab:landscape} by at most $0.010$, see \cref{sec:meth-noise}) and the fraction retained at $\sigma=0.3$ under geometry-respecting noise; the trained mean firing rate (three-seed probe, where measured); the normalised input-to-score sensitivity at probe scale $0.05$ (lower means scores more decoupled from input); and the Kendall rank correlation $\tau$ between clean and perturbed score orderings (higher means a more stable ranking). Every classical-spiking control retains $0.74$--$0.85$; the \qsvae alone retains $0.95$, at a ${\sim}3.5\times$ lower sensitivity and a far more stable ranking. Source: \texttt{runs/hep\_robustness\_whiten\_e\{5,6,7,8\}}, \texttt{runs/hep\_stress\_probes}, \texttt{runs/hep\_mechanism}.}
  \label{tab:supp-controls}
  \begin{tabular}{l c c c c c}
    \toprule
    Variant & Clean \auroc & Retention$@0.3$ & Firing & Sensitivity & Kendall $\tau$ \\
    \midrule
    \textbf{\qsvae} ($\sin^2$) & 0.830 & 0.950 & 0.23 & 0.17 & 0.83 \\
    FSVAE (LIF) & 0.868 & 0.795 & 0.29 & 0.64 & 0.48 \\
    Clamped, bn 1.0 & 0.871 & 0.839 & 0.37 & 0.62 & 0.43 \\
    Clamped, bn 0.80 & 0.870 & 0.831 & -- & 0.61 & 0.46 \\
    Clamped, bn 0.60 & 0.875 & 0.787 & -- & 0.64 & 0.49 \\
    Clamped, bn 0.45 (matched) & 0.877 & 0.765 & 0.19 & 0.70 & 0.46 \\
    Clamped, bn 0.30 & 0.882 & 0.741 & -- & 0.68 & 0.51 \\
    Logistic, $g=4$ & 0.867 & 0.847 & 0.37 & 0.63 & 0.42 \\
    Logistic, $g=2$ (slope-matched) & 0.863 & 0.849 & -- & 0.63 & 0.40 \\
    Logistic, $g=4$ (firing-matched) & 0.874 & 0.780 & -- & 0.66 & 0.45 \\
    FSVAE, wd $10^{-2}$ & 0.871 & 0.804 & -- & 0.61 & 0.48 \\
    FSVAE, wd $3{\times}10^{-2}$ & 0.867 & 0.799 & -- & 0.60 & 0.49 \\
    FSVAE, wd $10^{-1}$ & 0.875 & 0.809 & -- & 0.63 & 0.48 \\
    \bottomrule
  \end{tabular}
\end{table*}

\begin{table*}[!htb]
  \centering
  \caption{\textbf{Per-signal retention at $\sigma=0.3$ ($\ell=16$).} The fraction of clean \auroc retained under geometry-respecting noise, computed separately against each signal (three-seed mean for the spiking variants). The \qsvae's breakout is not an artefact of pooling: it holds on each signal, with the caveat that the broad-graviton column sits on a near-chance clean \auroc for the \qsvae and so carries little weight. Source: \texttt{runs/hep\_stress\_probes/persignal.csv}.}
  \label{tab:supp-retention-persignal}
  \begin{tabular}{l c c c c}
    \toprule
    Variant & Overall & Narrow & Broad & $A\!\to\!HZ$ \\
    \midrule
    \textbf{\qsvae} & 0.943 & 0.965 & 0.972 & 0.904 \\
    FSVAE & 0.814 & 0.878 & 0.804 & 0.756 \\
    Clamped & 0.848 & 0.914 & 0.822 & 0.798 \\
    Clamped (matched) & 0.786 & 0.846 & 0.786 & 0.725 \\
    Logistic, $g=4$ & 0.851 & 0.913 & 0.828 & 0.802 \\
    Logistic, $g=2$ & 0.854 & 0.913 & 0.840 & 0.803 \\
    Logistic (firing-matched) & 0.800 & 0.862 & 0.790 & 0.744 \\
    Isolation Forest & 0.961 & 0.985 & 0.962 & 0.935 \\
    \bottomrule
  \end{tabular}
\end{table*}

\begin{table*}[!htb]
  \centering
  \caption{\textbf{Overall \auroc across the full noise sweep ($\ell=16$, geometry-respecting).} The \qsvae's separation from every classical-spiking control opens from $\sigma\approx0.1$ and grows monotonically; the capstone's $\sigma=0.3$ is the endpoint of a trend, not a single-point artefact. Source: \texttt{runs/hep\_robustness\_whiten\_e\{5,6,7,8\}}.}
  \label{tab:supp-sigma-sweep}
  \begin{tabular}{l c c c c c c }
    \toprule
    Variant & $\sigma=0$ & $\sigma=0.05$ & $\sigma=0.1$ & $\sigma=0.15$ & $\sigma=0.2$ & $\sigma=0.3$ \\
    \midrule
    \textbf{\qsvae} ($\sin^2$) & 0.830 & 0.829 & 0.826 & 0.823 & 0.816 & 0.788 \\
    FSVAE (LIF) & 0.868 & 0.867 & 0.847 & 0.812 & 0.784 & 0.690 \\
    Clamped, bn 1.0 & 0.871 & 0.864 & 0.849 & 0.827 & 0.810 & 0.730 \\
    Clamped, bn 0.80 & 0.870 & 0.869 & 0.855 & 0.830 & 0.810 & 0.724 \\
    Clamped, bn 0.60 & 0.875 & 0.869 & 0.853 & 0.818 & 0.789 & 0.689 \\
    Clamped, bn 0.45 (matched) & 0.877 & 0.868 & 0.847 & 0.807 & 0.776 & 0.671 \\
    Clamped, bn 0.30 & 0.882 & 0.873 & 0.852 & 0.805 & 0.769 & 0.653 \\
    Logistic, $g=4$ & 0.867 & 0.866 & 0.854 & 0.828 & 0.812 & 0.735 \\
    Logistic, $g=2$ (slope-matched) & 0.863 & 0.863 & 0.851 & 0.827 & 0.810 & 0.733 \\
    Logistic, $g=4$ (firing-matched) & 0.874 & 0.867 & 0.850 & 0.812 & 0.782 & 0.682 \\
    FSVAE, wd $10^{-2}$ & 0.871 & 0.867 & 0.849 & 0.816 & 0.791 & 0.700 \\
    FSVAE, wd $3{\times}10^{-2}$ & 0.867 & 0.862 & 0.846 & 0.813 & 0.787 & 0.693 \\
    FSVAE, wd $10^{-1}$ & 0.875 & 0.870 & 0.853 & 0.818 & 0.797 & 0.708 \\
    \bottomrule
  \end{tabular}
\end{table*}

\begin{table}[!htb]
  \centering
  \caption{\textbf{Search metrics under the fair perturbation ($\ell=16$, $\sigma=0.3$, Overall).} Inverse background efficiency $1/\varepsilon_b$ at signal efficiency $\varepsilon_s=0.8$ and significance improvement $\mathrm{SIC}=\varepsilon_s/\sqrt{\varepsilon_b}$ at $\varepsilon_b=10^{-3}$, computed on the geometry-respecting noised test scores (three-seed mean). Under the perturbation the \qsvae leads every probed classical-spiking control on $1/\varepsilon_b$ and is level with the best of them on $\mathrm{SIC}$; the isolation forest still leads outright. Source: \texttt{runs/hep\_stress\_probes/persignal.csv}.}
  \label{tab:supp-search-noise}
  \begin{tabular}{l c c}
    \toprule
    Variant & $1/\varepsilon_b@0.8$ & $\mathrm{SIC}@10^{-3}$ \\
    \midrule
    \textbf{\qsvae} ($\sin^2$) & 2.1 & 2.5 \\
    FSVAE (LIF) & 1.7 & 1.5 \\
    Clamped, bn 1.0 & 1.9 & 2.4 \\
    Clamped, bn 0.80 & 1.9 & 1.9 \\
    Clamped, bn 0.60 & 1.8 & 1.2 \\
    Clamped, bn 0.45 (matched) & 1.7 & 0.9 \\
    Clamped, bn 0.30 & 1.6 & 0.7 \\
    Logistic, $g=4$ & 1.9 & 2.2 \\
    Logistic, $g=2$ (slope-matched) & 1.9 & 2.2 \\
    Logistic, $g=4$ (firing-matched) & 1.7 & 1.3 \\
    FSVAE, wd $10^{-2}$ & 1.8 & 1.5 \\
    FSVAE, wd $3{\times}10^{-2}$ & 1.7 & 1.4 \\
    FSVAE, wd $10^{-1}$ & 1.8 & 1.7 \\
    \bottomrule
  \end{tabular}
\end{table}

\bibliography{references}

\end{document}